\newdimen\tableauside\tableauside=1.0ex
\newdimen\tableaurule\tableaurule=0.4pt
\newdimen\tableaustep
\def\phantomhrule#1{\hbox{\vbox to0pt{\hrule height\tableaurule
width#1\vss}}}
\def\phantomvrule#1{\vbox{\hbox to0pt{\vrule width\tableaurule
height#1\hss}}}
\def\sqr{\vbox{%
  \phantomhrule\tableaustep

\hbox{\phantomvrule\tableaustep\kern\tableaustep\phantomvrule\tableaustep}%
  \hbox{\vbox{\phantomhrule\tableauside}\kern-\tableaurule}}}
\def\squares#1{\hbox{\count0=#1\noindent\loop\sqr
  \advance\count0 by-1 \ifnum\count0>0\repeat}}
\def\tableau#1{\vcenter{\offinterlineskip
  \tableaustep=\tableauside\advance\tableaustep by-\tableaurule
  \kern\normallineskip\hbox
    {\kern\normallineskip\vbox
      {\gettableau#1 0 }%
     \kern\normallineskip\kern\tableaurule}%
  \kern\normallineskip\kern\tableaurule}}
\def\gettableau#1 {\ifnum#1=0\let\next=\null\else
  \squares{#1}\let\next=\gettableau\fi\next}

\tableauside=1.0ex
\tableaurule=0.4pt

\newif\iflanl
\openin 1 lanlmac
\ifeof 1 \lanlfalse \else \lanltrue \fi
\closein 1
\iflanl
    \input lanlmac
\else
    \message{[lanlmac not found - use harvmac instead}
    \input harvmac
    \fi
\newif\ifhypertex
\ifx\hyperdef\UnDeFiNeD
    \hypertexfalse
    \message{[HYPERTEX MODE OFF}
    \def\hyperref#1#2#3#4{#4}
    \def\hyperdef#1#2#3#4{#4}
    \def\hypernoname{}
    \def\e@tf@ur#1{}
    \def\eprt#1{{\tt #1}}
    \def\CERN{\centerline{CERN, CH--1211 Geneva 23, Switzerland}}
    \def\wl{W.\ Lerche}
\else
    \hypertextrue
    \message{[HYPERTEX MODE ON}
\def\eprt#1{{\tt
#1}}
\def\CERN{\centerline{

$^a$Theory Division, CERN, Geneva, Switzerland}}
\def\wl{
 W.\ Lerche}
\fi
\newif\ifdraft

\noblackbox
\catcode`\@=11
\newif\iffrontpage
\ifx\answ\bigans
\def\titleft{\titla}
\magnification=1200\baselineskip=14pt plus 2pt minus 1pt
%
\advance\hoffset by-0.075truein
\advance\voffset by1.truecm
\hsize=6.15truein\vsize=600.truept\hsbody=\hsize\hstitle=\hsize
\else\let\lr=L
\def\titleft{\titla}
\magnification=1000\baselineskip=14pt plus 2pt minus 1pt
%
\hoffset=-0.75truein\voffset=-.0truein
\vsize=6.5truein
\hstitle=8.truein\hsbody=4.75truein
\fullhsize=10truein\hsize=\hsbody
\fi
\parskip=4pt plus 15pt minus 1pt
%
\newif\iffigureexists
\newif\ifepsfloaded
\def\epsfcheck{
\ifdraft
\input epsf\epsfloadedtrue
\else
  \openin 1 epsf
  \ifeof 1 \epsfloadedfalse \else \epsfloadedtrue \fi
  \closein 1
  \ifepsfloaded
    \input epsf
  \else
\immediate\write20{NO EPSF FILE --- FIGURES WILL BE IGNORED}
  \fi
\fi
\def\epsfcheck{}}
\def\checkex#1{
\ifdraft
\figureexistsfalse\immediate%
\write20{Draftmode: figure #1 not included}
\figureexiststrue
\else\relax
    \ifepsfloaded \openin 1 #1
        \ifeof 1
           \figureexistsfalse
  \immediate\write20{FIGURE FILE #1 NOT FOUND}
        \else \figureexiststrue
        \fi \closein 1
    \else \figureexistsfalse
    \fi
\fi}
\def\missbox#1#2{$\vcenter{\hrule
\hbox{\vrule height#1\kern1.truein
\raise.5truein\hbox{#2} \kern1.truein \vrule} \hrule}$}
\def\lfig#1{
\let\labelflag=#1%
\def\numb@rone{#1}%
\ifx\labelflag\UnDeFiNeD%
{\xdef#1{\the\figno}%
\writedef{#1\leftbracket{\the\figno}}%
\global\advance\figno by1%
}\fi{\hyperref{}{figure}{{\numb@rone}}{Fig.{\numb@rone}}}}
\def\figinsert#1#2#3#4{
\epsfcheck\checkex{#4}%
\def\figsize{#3}%
\let\flag=#1\ifx\flag\UnDeFiNeD
{\xdef#1{\the\figno}%
\writedef{#1\leftbracket{\the\figno}}%
\global\advance\figno by1%
}\fi
\goodbreak\midinsert%
\iffigureexists
\centerline{\epsfysize\figsize\epsfbox{#4}}%
\else%
\vskip.05truein
  \ifepsfloaded
  \ifdraft
  \centerline{\missbox\figsize{Draftmode: #4 not included}}%
  \else
  \centerline{\missbox\figsize{#4 not found}}
  \fi
  \else
  \centerline{\missbox\figsize{epsf.tex not found}}
  \fi
\vskip.05truein
\fi%
{\smallskip%
\leftskip 4pc \rightskip 4pc%
\noindent\ninepoint\sl \baselineskip=11pt%
{\bf{\hyperdef\hypernoname{figure}{{#1}}{Fig.{#1}}}:~}#2%
\smallskip}\bigskip\endinsert%
}

\def\boxit#1{\vbox{\hrule\hbox{\vrule\kern8pt
\vbox{\hbox{\kern8pt}\hbox{\vbox{#1}}\hbox{\kern8pt}}
\kern8pt\vrule}\hrule}}
\def\mathboxit#1{\vbox{\hrule\hbox{\vrule\kern8pt\vbox{\kern8pt
\hbox{$\displaystyle #1$}\kern8pt}\kern8pt\vrule}\hrule}}
%
\font\bigit=cmti10 scaled \magstep1

\font\titla=cmr10 scaled\magstep3
\font\tenmss=cmss10
\font\absmss=cmss10 scaled\magstep1

\newfam\mssfam
\font\footrm=cmr8  \font\footrms=cmr5
\font\footrmss=cmr5   \font\footi=cmmi8
\font\footis=cmmi5   \font\footiss=cmmi5
\font\footsy=cmsy8   \font\footsys=cmsy5
\font\footsyss=cmsy5   \font\footbf=cmbx8
\font\footmss=cmss8
\def\footfont{\def\rm{\fam0\footrm}
\textfont0=\footrm \scriptfont0=\footrms
\scriptscriptfont0=\footrmss
\textfont1=\footi \scriptfont1=\footis
\scriptscriptfont1=\footiss
\textfont2=\footsy \scriptfont2=\footsys
\scriptscriptfont2=\footsyss
\textfont\itfam=\footi \def\it{\fam\itfam\footi}
\textfont\mssfam=\footmss \def\mss{\fam\mssfam\footmss}
\textfont\bffam=\footbf \def\bf{\fam\bffam\footbf} \rm}
\def\tenpoint{\def\rm{\fam0\tenrm}
\textfont0=\tenrm \scriptfont0=\sevenrm
\scriptscriptfont0=\fiverm
\textfont1=\teni  \scriptfont1=\seveni
\scriptscriptfont1=\fivei
\textfont2=\tensy \scriptfont2=\sevensy
\scriptscriptfont2=\fivesy
\textfont\itfam=\tenit \def\it{\fam\itfam\tenit}
\textfont\mssfam=\tenmss \def\mss{\fam\mssfam\tenmss}
\textfont\bffam=\tenbf \def\bf{\fam\bffam\tenbf} \rm}
\ifx\answ\bigans\def\abstractfont{\tenpoint}\else
\def\abstractfont{\def\rm{\fam0\absrm}
\textfont0=\absrm \scriptfont0=\absrms
\scriptscriptfont0=\absrmss
\textfont1=\absi \scriptfont1=\absis
\scriptscriptfont1=\absiss
\textfont2=\abssy \scriptfont2=\abssys
\scriptscriptfont2=\abssyss
\textfont\itfam=\bigit \def\it{\fam\itfam\bigit}
\textfont\mssfam=\absmss \def\mss{\fam\mssfam\absmss}
\textfont\bffam=\absbf \def\bf{\fam\bffam\absbf}\rm}\fi
%
\def\f@@t{\baselineskip10pt\lineskip0pt\lineskiplimit0pt
\bgroup\aftergroup\@foot\let\next}
\setbox\strutbox=\hbox{\vrule height 8.pt depth 3.5pt width\z@}
\def\vfootnote#1{\insert\footins\bgroup
\baselineskip10pt\footfont
\interlinepenalty=\interfootnotelinepenalty
\floatingpenalty=20000
\splittopskip=\ht\strutbox \boxmaxdepth=\dp\strutbox
\leftskip=24pt \rightskip=\z@skip
\parindent=12pt \parfillskip=0pt plus 1fil
\spaceskip=\z@skip \xspaceskip=\z@skip
\Textindent{$#1$}\footstrut\futurelet\next\fo@t}
\def\Textindent#1{\noindent\llap{#1\enspace}\ignorespaces}
\def\foot{\global\advance\ftno by1%
\attach{\hyperref{}{footnote}{\the\ftno}{\footsymbolgen}}%
\vfootnote{\hyperdef\hypernoname{footnote}{\the\ftno}{\footsymbol}}}%
\def\footnote#1{\global\advance\ftno by1%
\attach{\hyperref{}{footnote}{\the\ftno}{#1}}%
\vfootnote{\hyperdef\hypernoname{footnote}{\the\ftno}{#1}}}%
\newcount\lastf@@t           \lastf@@t=-1
\newcount\footsymbolcount    \footsymbolcount=0
\global\newcount\ftno \global\ftno=0
\def\footsymbolgen{\relax\footsym
\global\lastf@@t=\pageno\footsymbol}
\def\footsym{\ifnum\footsymbolcount<0
\global\footsymbolcount=0\fi
{\iffrontpage \else \advance\lastf@@t by 1 \fi
\ifnum\lastf@@t<\pageno \global\footsymbolcount=0
\else \global\advance\footsymbolcount by 1 \fi }
\ifcase\footsymbolcount
\fd@f\dagger\or \fd@f\diamond\or \fd@f\ddagger\or
\fd@f\natural\or \fd@f\ast\or \fd@f\bullet\or
\fd@f\star\or \fd@f\nabla\else \fd@f\dagger
\global\footsymbolcount=0 \fi }
\def\fd@f#1{\xdef\footsymbol{#1}}
\def\space@ver#1{\let\@sf=\empty \ifmmode #1\else \ifhmode
\edef\@sf{\spacefactor=\the\spacefactor}
\unskip${}#1$\relax\fi\fi}
\def\attach#1{\space@ver{\strut^{\mkern 2mu #1}}\@sf}
%
\newif\ifnref
\def\rrr#1#2{\relax\ifnref\nref#1{#2}\else\ref#1{#2}\fi}
\def\ldf#1#2{\begingroup\obeylines
\gdef#1{\rrr{#1}{#2}}\endgroup\unskip}
\def\nrf#1{\nreftrue{#1}\nreffalse}
\def\doubref#1#2{\refs{{#1},{#2}}}

\nreffalse
\def\refout{\listrefs}

\def\lref{\ldf}

\def\eqn#1{\xdef #1{(\noexpand\hyperref{}%
{equation}{\secsym\the\meqno}%
{\secsym\the\meqno})}\eqno(\hyperdef\hypernoname{equation}%
{\secsym\the\meqno}{\secsym\the\meqno})\eqlabeL#1%
\writedef{#1\leftbracket#1}\global\advance\meqno by1}
\def\eqnalign#1{\xdef #1{\noexpand\hyperref{}{equation}%
{\secsym\the\meqno}{(\secsym\the\meqno)}}%
\writedef{#1\leftbracket#1}%
\hyperdef\hypernoname{equation}%
{\secsym\the\meqno}{\e@tf@ur#1}\eqlabeL{#1}%
\global\advance\meqno by1}
\def\eqnalign#1{\xdef #1{(\secsym\the\meqno)}
\writedef{#1\leftbracket#1}%
\global\advance\meqno by1 #1\eqlabeL{#1}}
%

%
\def\chap#1{\newsec{#1}}
\def\chapter#1{\chap{#1}}
\def\sect#1{\subsec{#1}}
\def\section#1{\sect{#1}}
\def\\{\ifnum\lastpenalty=-10000\relax
\else\hfil\penalty-10000\fi\ignorespaces}
\def\note#1{\leavevmode%
\edef\@@marginsf{\spacefactor=\the\spacefactor\relax}%
\ifdraft\strut\vadjust{%
\hbox to0pt{\hskip\hsize%
\ifx\answ\bigans\hskip.1in\else\hskip .1in\fi%
\vbox to0pt{\vskip-\dp
\strutbox\sevenbf\baselineskip=8pt plus 1pt minus 1pt%
\ifx\answ\bigans\hsize=.7in\else\hsize=.35in\fi%
\tolerance=5000 \hbadness=5000%
\leftskip=0pt \rightskip=0pt \everypar={}%
\raggedright\parskip=0pt \parindent=0pt%
\vskip-\ht\strutbox\noindent\strut#1\par%
\vss}\hss}}\fi\@@marginsf\kern-.01cm}
\def\titlepage{%
\frontpagetrue\nopagenumbers\abstractfont%
\hsize=\hstitle\rightline{\vbox{\baselineskip=10pt%
{\abstractfont\pubnum}}}\pageno=0}
\frontpagefalse
\def\pubnum{}
\def\pdate{\number\month/\number\yearltd}
\def\makefootline{\iffrontpage\vskip .27truein
\line{\the\footline}
\vskip -.1truein\leftline{\vbox{\baselineskip=10pt%
{\abstractfont\pdate}}}
\else\vskip.5cm\line{\hss \tenrm $-$ \folio\ $-$ \hss}\fi}
\def\title#1{\vskip .7truecm\titlestyle{\titleft #1}}
\def\titlestyle#1{\par\begingroup \interlinepenalty=9999
\leftskip=0.02\hsize plus 0.23\hsize minus 0.02\hsize
\rightskip=\leftskip \parfillskip=0pt
\hyphenpenalty=9000 \exhyphenpenalty=9000
\tolerance=9999 \pretolerance=9000
\spaceskip=0.333em \xspaceskip=0.5em
\noindent #1\par\endgroup }
\def\autskip{\ifx\answ\bigans\vskip.5truecm\else\vskip.1cm\fi}
\def\author#1{\vskip .7in \centerline{#1}}

\def\address#1{\ifx\answ\bigans\vskip.2truecm
\else\vskip.1cm\fi{\it \centerline{#1}}}
\def\abstract#1{
\vskip .5in\vfil\centerline
{\bf Abstract}\penalty1000
{{\smallskip\ifx\answ\bigans\leftskip 2pc \rightskip 2pc
\else\leftskip 5pc \rightskip 5pc\fi
\noindent\abstractfont \baselineskip=12pt
{#1} \smallskip}}
\penalty-1000}
\def\endpage{\tenpoint\supereject\global\hsize=\hsbody%
\frontpagefalse\footline={\hss\tenrm\folio\hss}}
\def\ack{\vskip2.cm\centerline{{\bf Acknowledgements}}}
%
%

%
\def\bfone{\relax{\rm 1\kern-.35em 1}}
\def\inbar{\vrule height1.5ex width.4pt depth0pt}
\def\IC{\relax\,\hbox{$\inbar\kern-.3em{\mss C}$}}
\def\ID{\relax{\rm I\kern-.18em D}}
\def\IF{\relax{\rm I\kern-.18em F}}
\def\IH{\relax{\rm I\kern-.18em H}}
\def\II{\relax{\rm I\kern-.17em I}}
\def\IN{\relax{\rm I\kern-.18em N}}
\def\IP{\relax{\rm I\kern-.18em P}}
\def\IQ{\relax\,\hbox{$\inbar\kern-.3em{\rm Q}$}}
\def\IR{\relax{\rm I\kern-.18em R}}
\font\cmss=cmss10 \font\cmsss=cmss10 at 7pt
\def\ZZ{\relax\ifmmode\mathchoice
{\hbox{\cmss Z\kern-.4em Z}}{\hbox{\cmss Z\kern-.4em Z}}
{\lower.9pt\hbox{\cmsss Z\kern-.4em Z}}
{\lower1.2pt\hbox{\cmsss Z\kern-.4em Z}}\else{\cmss Z\kern-.4em
Z}\fi}
\def\a{\alpha}

\def\cN{{\cal N}} 
 
\def\cR{{\cal R}} 
\def\nup#1({Nucl.\ Phys.\ $\us {B#1}$\ (}
\def\plt#1({Phys.\ Lett.\ $\us  {#1}$\ (}
\def\cmp#1({Comm.\ Math.\ Phys.\ $\us  {#1}$\ (}
\def\prp#1({Phys.\ Rep.\ $\us  {#1}$\ (}
\def\prl#1({Phys.\ Rev.\ Lett.\ $\us  {#1}$\ (}
\def\prv#1({Phys.\ Rev.\ $\us  {#1}$\ (}
\def\mpl#1({Mod.\ Phys.\ Let.\ $\us  {A#1}$\ (}
\def\ijmp#1({Int.\ J.\ Mod.\ Phys.\ $\us{A#1}$\ (}
\def\tit#1|{{\it #1},\ }
%

%

\def\ni{\noindent}
\def\tilde{\widetilde}
\def\bar{\overline}
\def\us#1{\underline{#1}}

\def\hat{\widehat}

\def\Coe#1.#2.{{#1\over #2}}
\def\coeff#1#2{\relax{\textstyle {#1 \over #2}}\displaystyle}
\def\coe#1.#2.{\relax{\textstyle {#1 \over #2}}\displaystyle}

\def\shalf{\relax{\textstyle {1 \over 2}}\displaystyle}

\def\to{\rightarrow}
\def\notin{\hbox{{$\in$}\kern-.51em\hbox{/}}}

\def\Trbel#1{\mathop{{\rm Tr}}_{#1}}

\def\del{\partial}

\def\nex#1{$N\!=\!#1$}


\def\ie{{\it i.e.}}

\catcode`\@=12


\def\LG{Lan\-dau-Ginz\-burg\ }
\def\N{{\cN}}
\def\nex#1{$\N\!=\!#1$}

\def\bfone{{\bf 1}}

\def\IG{\relax\,\hbox{$\inbar\kern-.3em{\mss G}$}}

\def\ga#1{\sigma_{#1}}
\def\cph#1#2{\widehat{CP}_{#1}^{\lower2pt\hbox{$\scriptstyle(#2)$}}}
\def\rnk#1#2{\cR_x^{(#1,#2)}}

\def\wnk#1#2{W^{(#1,#2)}}
\def\cph#1#2{{\rm CP}_{\!#1,#2}^{{\rm top}}}


\def\grl#1#2{^{\,[#1,#2]}}
\def\gkn#1#2{{\rm Gr}({#1}\!,{#2}\!)}
\def\Hst{\!H^*_{{\bar\del}}}
\def\ga(#1){{g_{(#1)}}}
\def\Fp{a}
\def\npk{{n\!+\!k}}
\def\npko{{n\!+\!k\!+\!1}}
\def\I{{\cal I}}
\def\Ih{{\hat \I}}
\def\wnk#1#2{W\grl{#1}{#2}}
\def\rnk#1#2{\cR\grl{#1}{#2}}
\def\aa(#1){\!\!{\fiverm#1}\!}
 1


\def\dim{{\rm dim}}

\def\frac#1#2{{#1\over #2}}

\def\zet{\ZZ}

\def\ee{{\rm e}}
\def\ii{{\rm i}}
\def\text#1{{\rm #1 }}

\def\N{{\cal N}}

\def\lieg{{\rm G}}
\def\lieh{{\rm H}}
\def\weyl{{w}}
\def\Weyl{{\rm W}}
\def\relweylgroup{{\Weyl(G)/\Weyl(H)}}
\def\weylsign{{\rm sign}(\weyl)}
\def\Ncos{{}^{\rm coset} N}
\def\NG{{}^G\! N}
\def\NH{{}^H\! N}

    \def\jw{J.\ Walcher}
\def\appA{A}
\def\appB{B}
\def\appC{C}



\def\nihil#1{{\sl #1}}
\def\br{\hfill\break}

\def\ijmp {{Int. J. Mod. Phys.\ }{\bf A}}

\lref\arn{V.\ Arnold, S.\ Gusein-Zade and A.\ Varchenko,
{\it Singularities of Differentiable Maps},
Birkh\"auser, 1988.}

\lref\MDcat{M.\ R.\ Douglas, 
\nihil{D-branes, Categories and N=1 Supersymmetry,}
\eprt{hep-th/0011017}. 
}

\lref\zub{J.\ Zuber, 
\nihil{CFT, BCFT, ADE and all that,}
\eprt{hep-th/0006151}. 
}

\lref\DGM{M.\ R.\ Douglas, B.\ R.\ Greene and D.\ R.\ Morrison, 
\nihil{Orbifold resolution by D-branes,}
 Nucl.\ Phys.\ {\bf B506} 84 (1997), 
\eprt{hep-th/9704151}. 
}

\lref\SCCVis{S.\ Cecotti and C.\ Vafa, 
\nihil{Ising model and N=2 supersymmetric theories,}
 Commun.\ Math.\ Phys.\ {\bf 157} 139 (1993), 
\eprt{hep-th/9209085}. 
}

\lref\GJ{S.\ Govindarajan and T.\ Jayaraman, 
\nihil{D-branes, exceptional sheaves and 
quivers on Calabi-Yau manifolds: From Mukai to McKay,}
\eprt{hep-th/0010196}. 
}

\lref\tomas{A.\ Tomasiello, 
\nihil{D-branes on Calabi-Yau manifolds and helices,}
\eprt{hep-th/0010217}. 
}

\lref\DDMD{D.\ Diaconescu and M.\ R.\ Douglas, 
\nihil{D-branes on stringy Calabi-Yau manifolds,}
\eprt{hep-th/0006224}. 
}

\lref\BDLR{I.\ Brunner, M.R.\ Douglas, A.\ Lawrence and 
C.\ R\"omelsberger, 
\nihil{D-branes on the quintic,}
\eprt{hep-th/9906200}. 
}

\lref\OOY{H.\ Ooguri, Y.\ Oz and Z.\ Yin, 
\nihil{D-branes on Calabi-Yau spaces and their mirrors,}
 Nucl.\ Phys.\ {\bf B477} 407 (1996), 
\eprt{hep-th/9606112}. 
}

\lref\bound{
See e.g.:
{A.\ Recknagel and V.\ Schomerus, 
\nihil{D-branes in Gepner models,}
 Nucl.\ Phys.\ {\bf B531} 185 (1998), 
\eprt{hep-th/9712186};\br 
}
{J.\ Fuchs and C.\ Schweigert, 
\nihil{Branes: From free fields to general backgrounds,}
 Nucl.\ Phys.\ {\bf B530} 99 (1998), 
\eprt{hep-th/9712257}. 
}
}

\lref\GovJay{S.\ Govindarajan and T.\ Jayaraman, 
\nihil{On the Landau-Ginzburg description of 
boundary CFTs and special Lagrangian submanifolds,}
 JHEP{\bf 0007} 016 (2000), 
\eprt{hep-th/0003242}. 
}

\lref\Bgep{
{S.\ Govindarajan, T.\ Jayaraman and T.\ Sarkar, 
\nihil{Worldsheet approaches to $D$-branes on supersymmetric cycles,}
 Nucl. Phys. {\bf B580} 519 (2000),           
\eprt{hep-th/9907131}; 
}
{D.\ Diaconescu and C.\ Romelsberger, 
\nihil{D-branes and bundles on elliptic fibrations,}
 Nucl.\ Phys.\ {\bf B574} 245 (2000), 
\eprt{hep-th/9910172}; 
}
\br
{P.\ Kaste, W.\ Lerche, C.\ A.\ L\"utken and J.\ Walcher, 
\nihil{$D$-branes on $K3$-fibrations,}
 Nucl.\ Phys.\ {\bf B582} 203 (2000), 
\eprt{hep-th/9912147}; 
}
\br
{E.\ Scheidegger, 
\nihil{D-branes on some one- and two-parameter Calabi-Yau hypersurfaces,}
 JHEP{\bf 0004} 003 (2000), 
\eprt{hep-th/9912188}; 
}
\br
{M.\ Naka and M.\ Nozaki, 
\nihil{Boundary states in Gepner models,}
 JHEP{\bf 0005} 027 (2000), 
\eprt{hep-th/0001037}; 
}
\br
{I.\ Brunner and V.\ Schomerus, 
\nihil{D-branes at singular curves of Calabi-Yau compactifications,}
 JHEP{\bf 0004} 020 (2000), 
\eprt{hep-th/0001132}; 
}
\br
{J.\ Fuchs, C.\ Schweigert and J.\ Walcher, 
\nihil{Projections in string theory and 
boundary states for Gepner models,}
 Nucl.\ Phys.\ {\bf B588} 110 (2000), 
\eprt{hep-th/0003298}; 
}
\br
{K.\ Sugiyama, 
\nihil{Comments on central charge of 
topological sigma model with Calabi-Yau target space,}
 Nucl.\ Phys.\ {\bf B591} 701 (2000), 
\eprt{hep-th/0003166}; 
}
\br
{W.\ Lerche, C.\ A.\ Lutken and C.\ Schweigert, 
\nihil{D-branes on ALE spaces and the 
ADE classification of conformal field theories,}
\eprt{hep-th/0006247}; 
}
\br
{W.\ Lerche, 
\nihil{On a boundary CFT description of 
nonperturbative N = 2 Yang-Mills theory,}
\eprt{hep-th/0006100}; 
}
\br
{J.\ Fuchs, P.\ Kaste, W.\ Lerche, 
C.\ A.\ Lutken, C.\ Schweigert and J.\ Walcher, 
\nihil{Boundary fixed points, enhanced gauge 
symmetry and singular bundles on K3,}
\eprt{hep-th/0007145}; 
}
\br
{I.\ Brunner and V.\ Schomerus, 
\nihil{On superpotentials for D-branes in Gepner models,}
 JHEP{\bf 0010} 016 (2000), 
\eprt{hep-th/0008194}; 
}
}

\lref\stanciu{S.\ Stanciu, 
\nihil{D-branes in Kazama-Suzuki models,}
 Nucl.\ Phys.\ {\bf B526} 295 (1998), 
\eprt{hep-th/9708166}. 
}

\lref\HIV{
{K.\ Hori and C.\ Vafa, 
\nihil{Mirror symmetry,}
\eprt{hep-th/0002222}; 
}\br
{K.\ Hori, A.\ Iqbal and C.\ Vafa, 
\nihil{D-branes and mirror symmetry,}
\eprt{hep-th/0005247}. 
}
}

\lref\OV{H.\ Ooguri and C.\ Vafa, 
\nihil{Two-Dimensional Black Hole and Singularities of CY Manifolds,}
 Nucl.\ Phys.\ {\bf B463} 55 (1996), 
\eprt{hep-th/9511164}. 
}

\lref\Kos{B.\ Kostant, 
\nihil{The principal three-dimensional subgroup and the
Betti numbers of a complex Lie group},
Am.\ J.\ Math.\ 81 (1959) 973.}

\lref\hill{
H.\ Hiller, \nihil{Geometry of Coxeter groups,}
Pitman London 1982; \br
J.\ Humphreys, 
\nihil{Reflection groups and Coxeter groups,}
Cambridge University Press 1990.}

\lref\LW{W.\ Lerche and N.\ P.\ Warner, 
\nihil{Polytopes and solitons in integrable, 
N=2 supersymmetric Landau-Ginzburg theories,}
 Nucl.\ Phys.\ {\bf B358} 571 (1991). 
}

\lref\cardy{J.\ L.\ Cardy, 
\nihil{Boundary Conditions, Fusion Rules And The Verlinde Formula,}
 Nucl.\ Phys.\ {\bf B324} 581 (1989). 
}

\lref\nickBLG{
{N.\ P.\ Warner, 
\nihil{Supersymmetry in boundary integrable models,}
 Nucl.\ Phys.\ {\bf B450} 663 (1995), 
\eprt{hep-th/9506064};  
}
{
\nihil{Supersymmetric, Integrable Boundary Field Theories,}
 Nucl.\ Phys.\ Proc.\ Suppl.\ {\bf 45A} 154 (1996), 
\eprt{hep-th/9512183}. 
}
}

\lref\EWgrass{E.\ Witten, 
\nihil{The Verlinde algebra and the cohomology of the Grassmannian,}
\eprt{hep-th/9312104}. 
}

\lref\FM{B.\ Fiol and M.\ Marino, 
\nihil{BPS states and algebras from quivers,}
 JHEP{\bf 0007} 031 (2000), 
\eprt{hep-th/0006189}. 
}

\lref\KS{
Y.~Kazama and H.~Suzuki,
\nihil{New N=2 Superconformal Field Theories And 
Superstring Compactification,}
Nucl.\ Phys.\  {\bf B321} (1989) 232.
}

\lref\LVW{
W.~Lerche, C.~Vafa and N.~P.~Warner,
\nihil{Chiral Rings In N=2 Superconformal Theories,}
Nucl.\ Phys.\  {\bf B324} (1989) 427.
}

\lref\FSC{
J.~Fuchs and C.~Schweigert,
\nihil{Symmetry breaking boundaries. I: General theory,
and II: More structures, examples,}
Nucl.\ Phys.\  {\bf B558} (1999) 419
\eprt{hep-th/9902132};
Nucl.\ Phys.\  {\bf B568} (2000) 543
\eprt{hep-th/9908025};
{
\nihil{Solitonic sectors, alpha-induction 
and symmetry breaking boundaries,}
\eprt{hep-th/0006181}. 
}
}

\lref\FLMW{P.\ Fendley, W.\ Lerche, S.\ D.\ Mathur and N.\ P.\ Warner, 
\nihil{$N=2$ supersymmetric integrable models from affine toda theories,}
 Nucl.\ Phys.\ {\bf B348} 66 (1991). 
}

\lref\FMVW{P.\ Fendley, S.\ D.\ Mathur, C.\ Vafa and N.\ P.\ Warner, 
\nihil{Integrable Deformations And 
Scattering Matrices For The $N=2$ Supersymmetric Discrete Series,}
 Phys.\ Lett.\ {\bf B243} 257 (1990). 
}

\lref\LW{W.\ Lerche and N.\ P.\ Warner, 
\nihil{Polytopes and solitons in integrable, 
N=2 supersymmetric Landau-Ginzburg theories,}
 Nucl.\ Phys.\ {\bf B358} 571 (1991). 
}

\lref\CVqr{C.\ Vafa, 
\nihil{Topological mirrors and quantum rings,}
\eprt{hep-th/9111017}. 
}

\lref\SCCVtt{S.\ Cecotti and C.\ Vafa, 
\nihil{Topological antitopological fusion,}
 Nucl.\ Phys.\ {\bf B367} 359 (1991). 
}

\lref\SCCVss{S.\ Cecotti and C.\ Vafa, 
\nihil{Exact results for supersymmetric sigma models,}
 Phys.\ Rev.\ Lett.\ {\bf 68} 903 (1992), 
\eprt{hep-th/9111016}. 
}

\lref\SCCVcn{S.\ Cecotti and C.\ Vafa, 
\nihil{On classification of N=2 supersymmetric theories,}
 Commun.\ Math.\ Phys.\ {\bf 158} 569 (1993), 
\eprt{hep-th/9211097}. 
}

\lref\KI{K.\ Intriligator, 
\nihil{Fusion residues,}
 Mod.\ Phys.\ Lett.\ {\bf A6} 3543 (1991), 
\eprt{hep-th/9108005}. 
}

\lref\FI{P.\ Fendley and K.\ Intriligator, 
\nihil{Scattering and thermodynamics in integrable N=2 theories,}
 Nucl.\ Phys.\ {\bf B380} 265 (1992), 
\eprt{hep-th/9202011}. 
}

\lref\EWqh{
{E.\ Witten, 
\nihil{Topological Sigma Models,}
 Commun.\ Math.\ Phys.\ {\bf 118} 411 (1988). 
};
{ 
\nihil{Two-dimensional gravity and intersection 
theory on moduli space,}
 Surveys Diff.\ Geom.\ {\bf 1} 243 (1991). 
}
}

\lref\DGfr{D.\ Gepner, 
\nihil{Fusion rings and geometry,}
 Commun.\ Math.\ Phys.\ {\bf 141} 381 (1991). 
}

\lref\DF{M.\ R.\ Douglas and B.\ Fiol, 
\nihil{D-branes and discrete torsion.\ II,}
\eprt{hep-th/9903031}. 
}

\lref\dFZ{P.\ Di Francesco and J.\ B.\ Zuber, 
\nihil{Fusion potentials.\ 1,}
 J.\ Phys.\ A{\bf A26} 1441 (1993), 
\eprt{hep-th/9211138}. 
}

\lref\JZg{J.\ B.\ Zuber, 
\nihil{Graphs and reflection groups,}
 Commun.\ Math.\ Phys.\ {\bf 179} 265 (1996), 
\eprt{hep-th/9507057}. 
}

\lref\FLZ{P.\ Di Francesco, F.\ Lesage and J.\ B.\ Zuber, 
\nihil{Graph rings and integrable 
perturbations of N=2 superconformal theories,}
 Nucl.\ Phys.\ {\bf B408} 600 (1993), 
\eprt{hep-th/9306018}. 
}

\lref\GZAV{S.\ M.\ Gusein-Zade and A.\ Varchenko, 
\nihil{Verlinde algebras and the 
intersection form on vanishing cycles,}
\eprt{hep-th/9610058}. 
}

\lref\Dgep{D.\ Gepner, 
\nihil{Space-Time Supersymmetry In Compactified
String Theory And Superconformal Models,} 
Nucl.\ Phys.\ {\bf B296}, 757 (1988).}

\lref\topw{
{W.\ Lerche and A.\ Sevrin, 
\nihil{On the Landau-Ginzburg realization of topological gravities,}
 Nucl.\ Phys.\ {\bf B428} 259 (1994), 
\eprt{hep-th/9403183};
}
\br
{W.\ Lerche and N.\ P.\ Warner, 
\nihil{On the algebraic structure of gravitational 
descendants in $CP^{n-1}$ coset models,}
 Phys.\ Lett.\ {\bf B343} 87 (1995), 
\eprt{hep-th/9409069}. 
}
}


\lref\LLS{W.\ Lerche, C.\ A.\ L\"utken and C.\ Schweigert, 
as in ref.\ \Bgep.}

\lref\bries{E.\ Brieskorn,
\nihil{The unfolding of exceptional singularities},
Nova acta Leopoldina NF 52 Nr.\ 240, (1981) 65-93.}

\lref\JSmc{Y.\ He and J.\ S.\ Song, 
\nihil{Of McKay correspondence, 
non-linear sigma-model and conformal field theory,}
\eprt{hep-th/9903056}. 
}

\lref\JSgor{J.\ S.\ Song, 
\nihil{Three-dimensional Gorenstein 
singularities and $\widehat SU(3)$ modular invariants,}
\eprt{hep-th/9908008}. 
}

\lref\WLwgrav{W.\ Lerche, 
\nihil{Generalized Drinfeld-Sokolov hierarchies, 
quantum rings and W gravity,}
 Nucl.\ Phys.\ {\bf B434} 445 (1995), 
\eprt{hep-th/9312188}. 
}

\lref\McKay{
{J.\ McKay, 
\nihil{Graphs, singularities and finite groups,}
Proc.\ Symp.\ Pure Math.\ {\bf 37} (1980) 183;
}
\br
{M.\ Reid, 
\nihil{McKay correspondence,}
 \eprt{alg-geom/9702016}. 
}
}

\lref\DFR{M.\ R.\ Douglas, B.\ Fiol and C.\ R\"omelsberger, 
\nihil{The spectrum of BPS branes on a noncompact Calabi-Yau,}
\eprt{hep-th/0003263}. 
}

\lref\KosA{B.\ Kostant, private communication.}

\lref\naka{H. Nakajima, 
\nihil{Instantons on ALE spaces, quiver varieties, 
and Kac-Moody algebras,}
Duke Math.~{\bf 76} (1994) 365;
\nihil{Gauge theory on resolutions of simple singularities
and affine Lie algebras,}
Int.\ Math.\ Res.\ Not.\ {\bf 2} (1994) 61;
\nihil{Instantons and affine Lie algebras,}
\eprt{alg-geom/9502013.}
}

\lref\PMMcMay{P.\ Mayr, 
\nihil{Phases of supersymmetric D-branes on 
K\"ahler manifolds and the McKay correspondence,}
\eprt{hep-th/0010223}. 
}

\lref\ITNA{
{Y.~Ito and M.~Reid, 
\nihil{The McKay correspondence for finite subgroups of $SL(3,\IC)$,} 
\eprt{math.AG/9411010};
}
\br
{Y.~Ito and H.~Nakajima, 
\nihil{McKay correspondence 
and Hilbert schemes in dimension three,} 
\eprt{math.AG/9803120}.
}
}

\lref\yhe{Y.\ He, 
\nihil{Some remarks on the finitude of quiver theories,}
\eprt{hep-th/9911114}. 
}

\lref\MDGM{M.\ R.\ Douglas and G.\ Moore, 
\nihil{D-branes, Quivers, and ALE Instantons,}
\eprt{hep-th/9603167}. 
}

\lref\bert{A.\ N.\ Schellekens, 
\nihil{Field identification fixed points in N=2 coset theories,}
 Nucl.\ Phys.\ {\bf B366} 27 (1991). 
}



\def\pubnum{
\hbox{CERN-TH/2000-335}
\hbox{ETH-TH/00-11}
\hbox{hep-th/0011107}
\hbox{}}
\def\pdate{}
\titlepage
\vskip2.cm
\title{{\titlefont 
Boundary Rings and $\cN\!\!=\!\!2$ Coset Models}}
\vskip -.7cm
\autskip
\author{
\wl$^a$\ and \jw$^{a,b}$}  
\vskip0.7truecm
\CERN
\vskip.2truecm
\centerline{$^b\,$Institut f\"ur Theoretische Physik, 
ETH-H\"onggerberg, CH-8093 Z\"urich, Switzerland} 

\abstract{
We investigate boundary states of \nex2\ coset models based on
Grassmannians  $\gkn n\npk$, and find that the underlying intersection
geometry is given by the fusion ring of $U(n)$. This is isomorphic to
the quantum cohomology ring of $\gkn n\npko$,  which in turn can be
encoded in a ``boundary'' superpotential  whose critical points
correspond to the boundary states. In this way the intersection
properties can be represented in terms of a soliton graph that forms a
generalized, $\ZZ_{n+k+1}$ symmetric McKay quiver. We investigate the
spectrum of bound states and find that the rational boundary CFT
produces only a small subset of the possible quiver representations.
}

\vfil
\ni {CERN-TH/2000-335}\hfill\break
\ni November 2000
\endpage
\baselineskip=14pt plus 2pt minus 1pt


\chapter{Introduction}

\nrf{\BDLR\DFR\HIV\DDMD\PMMcMay\MDcat} 
There has been exciting
recent progress in the understanding of the quantum geometry of
$D$-branes at both large and small volume of the compactification
Calabi-Yau manifold, see for example refs.~\refs{\BDLR--\MDcat}. At
large volume the theory is most naturally described in classical terms,
i.e., by $D$-branes wrapped on non-trivial cycles, which in
mathematical language corresponds to submanifolds with bundles or
sheaves on them. On the other hand, at small volume the quantum
corrections are strong so that the classical geometrical picture must
fail. One can nonetheless obtain exact results in the strong coupling
region by using a conformal field theory description in which the
$D$-branes are represented as boundary states. The CFT's that have been
successfully employed for this purpose include orbifolds
$\IC^n/\Gamma$ \doubref\MDGM\DGM\  and ``Gepner'' \Dgep\ tensor
products \doubref\BDLR\Bgep\ of the  \nex2 superconformal minimal models
\bound.  The boundary sector of such theories can be described in terms
of quiver theory \MDGM, which is the appropriate mathematical framework to
describe the physics of the small-volume phase. Very recently it was
shown \refs{\DDMD,\GJ,\tomas,\PMMcMay}  how the two descriptions at
large and small volume can be  related via a generalized McKay
correspondence \doubref\McKay\ITNA,  which gives a precise map between
the large radius bundle data and the quiver group theory data at small
radius.

However the set of exactly solvable rational CFT is much larger than
orbifolds and \nex2 minimal models, and tensor products of them. In
particular there are \nex2 superconformal field theories based on
cosets $SU(n+1)_k/U(n)$ \KS, which generalize the minimal models (for
which $n=1$). From the CFT point of view, these models are on a similar
footing as the minimal models, so that it is a natural question to ask
about the properties of the boundary states of these models. On the
other hand, from a geometrical point of view they correspond to
isolated singularities that are not necessarily of orbifold type, so we
may expect to find novel features  with regard to generalizations of
the McKay correspondence. Indeed these models have an abundantly rich
mathematical structure (related to Grassmannians $\gkn n{n+k}\cong
U(\npk)/U(n)\times U(k)$) \nrf{\LVW\DGfr\LW\KI\SCCVcn\EWgrass\FLZ} that
has been analyzed in great detail in the past (see e.g.,
\refs{\LVW--\FLZ}), as far as the bulk physics is concerned.

Our purpose is to make a first step in unraveling the properties of the
boundary sector, by focusing on the intrinsic, algebraic aspects of the
coset boundary CFT.\foot{See ref.\ \stanciu\ for some other aspects of
$D$-branes in Kazama-Suzuki models.}  We will investigate tensor
products of such models and their geometrical and $K$-theoretical
significance with respect to sheaves on Grassmannians in a subsequent
paper.

In the next section we will outline our general ideas for the \nex2
minimal models which are based on $SU(2)_k/U(1)$.  In particular we will
show that the natural algebraic structure underlying the boundary
states is given by a ``boundary fusion ring''.
\foot{Boundary rings (in a more geometrical setting) were first
introduced in \PMMcMay.} It is isomorphic  to the
representation ring of $\ZZ_{k+2}$ which figures in the McKay
correspondence for $\IC^2/\ZZ_{k+2}$ \McKay.  In the subsequent
sections, we will  first introduce the \nex2 coset models based on
$SU(n+1)_k/U(n)$ and then compute the boundary state intersection index
$\I_{a,b}\equiv\Trbel{a,b}[(-1)^F]_{RR}$. We will analyze its algebraic
structure in some detail, and find connections to previous work on the
quantum cohomology of Grassmannians, soliton polytopes, 2d gravity and
quiver theory.

Specifically we will find a boundary ring that is analogous to the one
of the minimal models, essentially given by the fusion ring of $U(n)$.
It can be viewed as the path algebra of the underlying quiver. While
we will not succeed to find a generalized McKay correspondence
\doubref\JSmc\JSgor\ involving discrete groups, we find a weaker McKay
correspondence   in that the intersection homology of the resolution of
the isolated singularity corresponding to the coset model
\doubref\JZg\GZAV\ is correctly reproduced by the boundary fusion ring.
We will also investigate the spectrum of bound states and find that the
present boundary CFT methods generate only a very small subset of all
the possible quiver representations. Some technical details will be
deferred to appendices.

\section{$A_{k+1}$ minimal models and the McKay quiver for ALE spaces}

As a warm-up, we will first illustrate our ideas  for the \nex2\
minimal models of type $A_{k+1}$ \KS. In the subsequent sections, we
will then discuss the more general class of Kazama-Suzkui coset models.

The minimal models are based on the cosets $SU(2)_{k}/U(1)$, or
equivalently on $SU(k\!+\!1)_1/ U(k)$. Each formulation has a particular
geometrical significance that we will elucidate in the following.
Specifically, $SU(2)_{k}$ is naturally tied to ALE spaces that resolve
$\IC^2/\ZZ_{k\!+\!2}$. On the other hand \doubref\LVW\DGfr\ the
structure of the chiral ring $\cR\grl1k\!=\!\{1,x,...,x^{k}\}$ can
be most naturally understood in terms of the ``level-1'' formulation,
in that it is isomorphic to the cohomology ring $\Hst(\IP^{k},\IR)$,
where $\IP^{k}\!\!=\! {SU(k\!+\!1)/U(k)}$.

If we use the $SU(2)_k$ formulation for the minimal models, the
primary fields are labelled by $(\ell,m,s)$, with $\ell=0,...,k$,
$m=-k-1,\dots,k+2$ (mod $2k\!+\!4$), and in addition $s=-1,0,1,2$ (mod
$4$) determines the R- or NS-sectors $(\ell+m+s=0$ mod $2$).  We will
be interested in boundary states $|\ell,m,s\rangle$, which are
labelled by the same letters as the primary fields.

In a given sector we thus have for each $\ell$ an orbit of $k\!+\!2$
boundary states. The states with $\ell=0$ can be viewed as the basic
states, out of which the higher spin states can be formed as bound
states. More precisely, to be able to specify a natural ordering,
say in terms of the masses of wrapped $D$-branes, we have to perturb
the theory such as to resolve the singularity and to provide a mass
scale. This procedure is not unique, but if we like to maintain the
$\ZZ_{k+2}$ ``Coxeter'' $R$-symmetry in the $m$-labels, 
we are lead to considering the following superpotential:
$$
W(x,\mu)\grl1k\ =\ x^{k+2}+\mu\ ,
\eqn\Wlambda
$$
where the constant $\mu$ sets the scale.
It may be viewed as the inhomogenous form of the Landau-Ginzburg
potential $W_{ALE}^{A_{k+1}}(x,z)=x^{k+2}+\mu z^{-k-2}$, which describes a
type II string compactification on a $\ZZ_{k+2}$-symmetrically resolved
non-compact ALE space\foot{To be precise one would need to add two
further quadratic terms in order to get the correct dimension of the
manifold, but these terms are not important for our purposes.} of type
$A_{k+1}$ \OV; see \lfig\AnDynkin.

\figinsert\AnDynkin{On the left we see the manifold $x^{k+2}+\mu=0$
consisting of $(k\!+\!2)$ points (here $k=3$). This should
be viewed as a model for the homology of an ALE space, where the
dashed lines denote $2$-cycles. On the right we see the critical
points of the perturbed LG potential  $x^{k+3}+\mu\,x$ which
correspond to the boundary states associated with the cycles on the left
figure. The links depict the fundamental solitons and form the
graph of the BCFT intersection matrix $\Ih_{0,0}$, given by the
extended Dynkin diagram $\hat A_{k+1}$. It is the McKay quiver
associated with the ALE resolution of $\IC^2/\ZZ_{k+2}$.}{1.in}{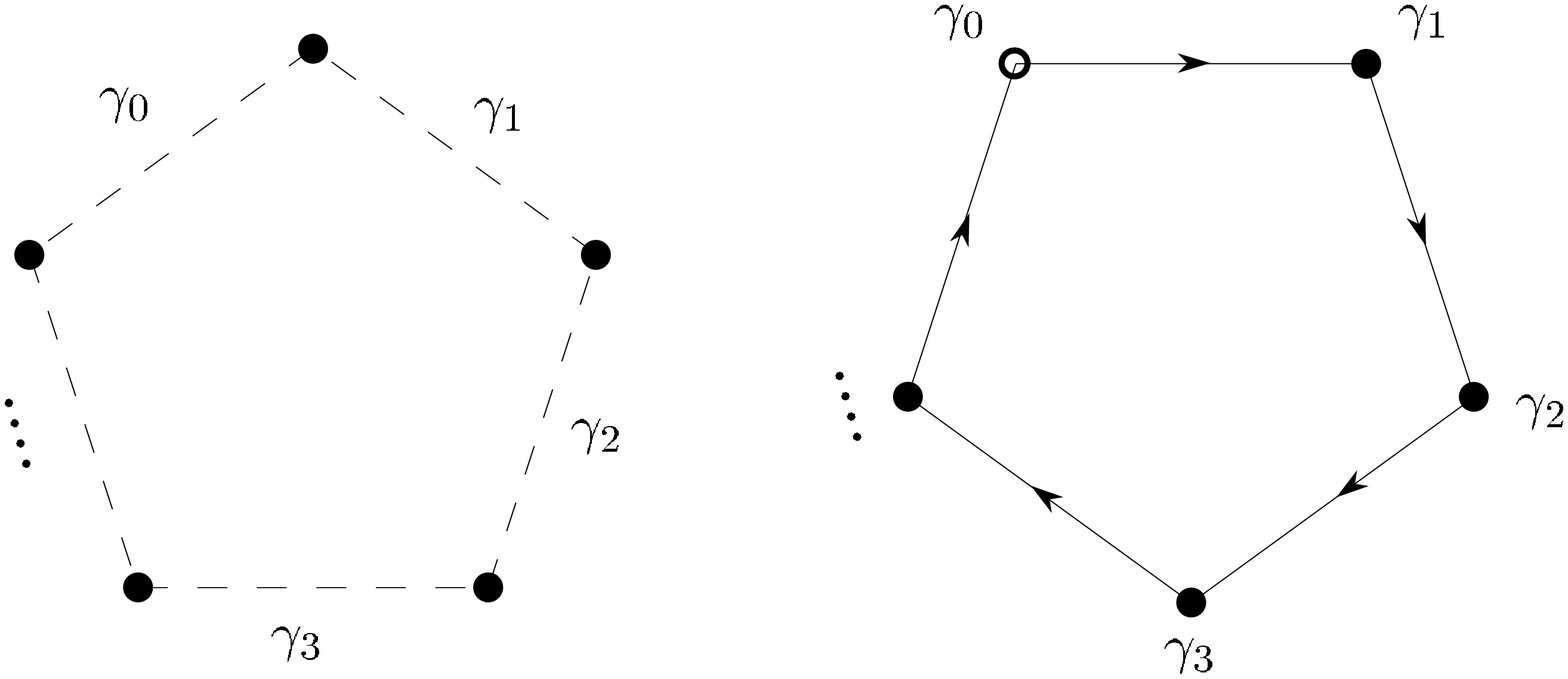}

The quantity of interest is the 
topological intersection index \DF\ $I_{a,b}\equiv 
{\rm Tr}_{a,b}[(-1)^F]$ between boundary states $a$,$b$,
which is independent of $\mu$.
As has been shown in recent papers \doubref\BDLR\HIV, 
it can be represented as an overlap amplitude:
$$
\Ih_{{\ell_1},{\ell_2}}(m_1,m_2,s_1,s_2)\ \equiv\
{\phantom{\big|}}_{{\rm RR}}\big\langle \ell_1,m_1,s_1\big|
\ell_2,m_2,s_2\big\rangle_{{\rm RR}}
\ .\eqn\defI
$$
Using the well-known expansion of the boundary states into Ishibashi states,
with the appropriate normalization, one gets as result \doubref\BDLR\HIV:
$$
\big(\Ih_{{\ell_1},{\ell_2}}\grl1{k}\big)_{m_1}^{\ \ m_2}(s_1,s_2)\ =\
(-1)^{{s_2-s_1\over2}}\,N_{{\ell_1},{\ell_2}}^{m_2-m_1}\ .
\eqn\IeqN
$$
It can be considered as a $(2k\!+\!4)\times(2k\!+\!4)$ matrix for fixed 
$\ell_i$, $s_i$ (in the following, we will keep $s_i$ fixed).
Here
$$
N_{{\ell_1},{\ell_2}}^{\ell_3}\! =
{2\over k\!+\!2}\sum_{\ell=0}^k
{
\sin\!\big[\!{\pi\over k\!+\!2}(\ell_1\!+\!1)(\ell\!+\!1)\big]
\sin\!\big[\!{\pi\over k\!+\!2}(\ell_2\!+\!1)(\ell\!+\!1)\big]
\sin\big[\!{\pi\over k\!+\!2}(\ell_3\!+\!1)(\ell\!+\!1)\big]
\over
\sin\big[{\pi\over k\!+\!2}(\ell\!+\!1)\big]
}\eqn\FR
$$
are nothing but the Verlinde fusion coefficients
associated with $SU(2)_k$. Note however that since $m_2$ can be smaller
than $m_1$ and moreover both labels are periodic, the range of the
upper index must be continued beyond the standard range of integrable
representations of $SU(2)_k$, by defining
$N_{{\ell_1},{\ell_2}}^{-\ell_3-2}\equiv
-N_{{\ell_1},{\ell_2}}^{\ell_3}$ and
$N_{{\ell_1},{\ell_2}}^{-1}=N_{{\ell_1},{\ell_2}}^{k-1}\equiv0$. In
effect \IeqN\ should rather be viewed in terms of the fusion
coefficients of $U(1)_{2k+4}$. These are given by powers
of the $\ZZ_{\!2k\!+\!4}$ step generator $\ga(\!2k\!+\!4)$, 
where
$$
\ga(M)\  :\equiv\  \left( 
\matrix{ 0 & 1 & 0 & \cdots & 0 & 0 \cr
         0 & 0 & 1 & \cdots & 0 & 0 \cr
         \vdots & \vdots & \vdots & \ddots & \vdots & \vdots \cr
         0 & 0 & 0 & \cdots & 0 & 1  \cr
     1 & 0 & 0 & \cdots & 0 & 0} \right)_{M\times M}\ . 
\eqn\shift
$$
More precisely, since due to the selection rule $l+m+s=0$ (mod $2$)
only even powers contribute for $\ell_1=\ell_2$, we effectively have a
smaller group, $U(1)_{k+2}$, and so we may express $\Ih_{\ell,\ell}$ more
efficiently  in terms of the reduced matrices $\ga(\!k\!+\!2)$. We can write in
particular for the basic $\ell=0$ states:
$$
\Ih\grl1{k}\ \equiv\ \Ih_{0,0}\grl1{k}\ 
=\ \sum_{p=0}^1 (-1)^p \hat\Fp\grl1{k}_p\ ,
\eqn\Inullnullext
$$ 
where  $\hat\Fp_p\grl1{k}\equiv (\ga(\!k\!+\!2))^p$ are the fusion
coefficients of $U(1)_{k+2}$ (the notation should become clear further
below; in particular, the hat denotes ``affine extension'').

The formula \Inullnullext\ has multiple geometrical 
and physical significance, which will be 
worked out for general Grassmannian coset models in the
subsequent sections. In the present context, 
we will restrict ourselves to the following remarks.

First of all, \Inullnullext\ yields upon symmetrization the
well-known intersection form of the homology of the ALE
space:\foot{The transposed contribution can be 
attributed to the non-compact piece
in $W_{ALE}^{A_{k+1}}$.}
$$
C_{\hat A_{k+1}}\ =\ \Ih\grl1{k}+(\Ih\grl1{k})^t
\ ,
\eqn\ALECartext
$$
where $C_{\hat A_{k+1}}$ is the {\it extended} Cartan matrix of
$A_{k+1}$. The reason why we get the extended Cartan matrix is because
the $\ZZ_{k\!+\!2}$ orbit of boundary states with $\ell=0$  corresponds
to an over-complete homology basis. A minimal choice for the homology
basis can be obtained by dropping the last row and column of
$\ga(\!k\!+\!2)$. The resulting upper-triangular matrices
$\Fp_p\grl1{k}$ are then precisely the  structure constants of the bulk
chiral ring, $\cR\grl1k\!=\!\{1,x,...,x^{k}\}$. These give rise to the
``reduced'' intersection form
$$
\I\grl1{k}\ =\ \sum_{p=0}^1 (-1)^p \Fp\grl1{k}_p\ ,
\eqn\Inullnull
$$ 
whose symmetrization gives the ordinary Cartan matrix:
$$
C_{A_{k+1}}\ =\ \I\grl1{k}+(\I\grl1{k})^t\ .
\eqn\ALECart
$$
It also coincides with the fusion ring structure constants of
$SU(2)_k$. Thus the Dynkin diagram of $A_{k+1}$ 
plays here a dual r\^ole as the fusion graph \JZg\ of $SU(2)_k$.

Returning to the extended intersection form \Inullnullext, we note that
it can be interpreted as a sum over intermediate
open string states $\varphi_{[p]}$ associated with the fundamental
representations $[p]$ of $U(1)_{k+2}$. These can be characterized by
Young tableaux with $p=0,1$ boxes, respectively; the other
representations $\rho_a$ have a single row with up to $k$ boxes. In
this way the two-point function $\Ih_{a}^{\ b}=\langle \bar
\rho_b|\sum (-)^p\varphi_{[p]}|\rho_a\rangle$ can be associated with an
operator algebra of the form:
$$
[p]\otimes \rho_a\ =\ \bigoplus_b (\hat\Fp\grl1{k}_p)_{ab}\rho_b\ .
\eqn\Fprule
$$
Upon restriction to the discrete subgroup  $\ZZ_{k+2}\subset
U(1)_{k+2}$ that has been left unbroken by the resolution, this formula
expresses the McKay correspondence \McKay\ associated with the ALE space
resolution of $\IC^2/\ZZ_{k+2}$. It relates the intersection homology
of  the resolution of the singularity $\IC^2/\ZZ_{k+2}$ to the
representation ring $\cR(\ZZ_{k+2})$.\foot{An analogous interpretation
can be made for the non-extended intersection matrix \Inullnull, where
the $U(1)$ fusion coefficients $\hat\Fp\grl1{k}_p$ in \Fprule\ get
replaced by the upper triangular, chiral ring structure constants
$\Fp_p\grl1{k}$.  The main difference to the affine extended situation
is that this is a nilpotent ring and not the representation ring of a
discrete group.}

In fact also the $U(1)_{k+2}$ fusion coefficients
$\hat\Fp\grl1{k}_p$ can formally be thought of as structure constants of a
chiral ring of some Landau-Ginzburg theory. 
Namely they are nothing but the chiral ring structure constants 
associated with the following  ``boundary fusion superpotential'': 
$$
\hat W(x,\mu)\grl1{k}\ =\ \coeff1{k+3} x^{k+3}+\mu\,x
\eqn\boundaryW
$$
at the special point $\mu=-1$ (the perturbation leads to the
non-vanishing lower left corner entry in the step generator matrix
$\hat\Fp_1=\ga(k+2)$ in \shift). The underlying mathematical reason is
the fact \EWgrass\ that the fusion ring of $U(1)_{k+2}$ is isomorphic
(for $\mu=-1$) to the quantum cohomology ring  of $\IP^{k+1}$, which  is
given by $\IC[x]/(x^{k+2}-\lambda)$ \refs{\EWqh,\KI,\SCCVss,\CVqr}.
The appearance of a potential $\hat W(x)$ with the property that its
derivative yields the bulk superpotential,
$$
W(x,\mu)\grl1{k}\ =\ {\del\over\del x}\hat W(x,\mu)\grl1{k}\ ,
\eqn\derivrel
$$
is natural in boundary LG theory --  it was shown \nickBLG\
that if one requires non-trivial boundary dynamics while maintaining
\nex2 supersymmetry, one needs to include $\hat W$ as a boundary
superpotential.

In physics terms this is similar to the findings of 
ref.~\doubref\GovJay\HIV, where
$D$-branes are associated with critical points of superpotentials. More
precisely, the critical points (minima of the potential $\hat
V=|\del\hat W|^2$, given here by the $(k+2)$-th roots of unity) are
linked by solitons which map to straight lines in the $\hat W$-plane
\doubref\FMVW\SCCVcn. According to \doubref\GovJay\HIV, the number of
solitons that map between a given pair of critical points reflects the
number of open string states stretching between $D$-branes, and hence
coincides with the intersection number of the $D$-branes.

However our approach is different as compared to \HIV, in that our
$\hat W$-plane does not describe central charges of LG solitons. Rather
it realizes the intersection graph of the $(\ell=0, m\in \ZZ_{k+2})$
orbit of boundary states, which correspond to the
critical points of $\hat W(x,\mu)\grl1{k}$. 
Indeed we explicitly see from \lfig\AnDynkin\ 
that the fundamental ``solitons'' of the perturbed
LG potential \boundaryW\ yield the graph of the $\ell=0$ boundary state
intersection form $\Ih\grl1{k}$, i.e., the affine Dynkin diagram of
$\hat A_{k+1}$; they correspond to the fermionic open string zero modes
$\varphi_{[1]}$.

While the topological soliton intersection numbers do not depend on the
precise value of $\mu$ (apart from monodromy), the special value
$\mu=-1$ is distinguished in that only at this value we can use the chiral ring
structure constants to conveniently compute the intersection index as in
\Inullnullext. This is similar to the results of \doubref\DGfr\dFZ,
where the fusion rules of $SU(n+1)_k$ were reproduced from the
perturbation of superpotentials by generalized Chebychev polynomials,
for a certain fixed value of the perturbing parameter~$\mu$.

\section{\nex2 coset models}

We will now recall some known facts about the Kazama-Suzuki models
\KS. These are rational \nex2\ superconformal field
theories defined by the coset construction.
A generic theory is denoted by
$$
\left(\frac{\lieg\times SO(2d)}{\lieh\times U(1)}\right)_k\ ,
\eqn\kasudef
$$
where $k$ is the level for the affine Lie algebra $\lieg$ and the
$SO(2d)$ factor arises from the fermions and is at level $1$.
Furthermore, $2d=\dim\lieg-\dim\lieh$. We will often use the notation
$\lieg_k/\lieh$ as a shorthand for \kasudef. Our main interest is in
models where $\lieg$ is simply laced  at level  one, and the
underlying coset space is a hermitian  symmetric space. More
specifically we will be interested in this paper in models based on
Grassmannians $\gkn n{n+k}$, for which the following equivalences hold:
$$
{SU(n+k)_1\over SU(n)\times SU(k)\times U(1)}
\cong 
{SU(n+1)_k\over SU(n)\times U(1)}
\cong
{SU(k+1)_n\over SU(k)\times U(1)}\ .
\eqn\KSdef
$$
We will label quantities pertaining to these models by
a superscript $[n,k]$, and we will assume as convention $n\leq k$.

The definition of the coset \kasudef\ includes the specification
of the embedding of $\lieh$ into $\lieg$, and is accompanied
by specific selection rules and field identifications. Field
identification fixed points do not occur for the models we consider, 
so we can neglect this complication here. Let us point out, however,
that since fixed point resolution affects the modular data
and fusion rules in a non-trivial way, it will have interesting
consequences for the intersection index of boundary states
in theories with fixed points.

Primary (with respect to the bosonic algebra) fields in the coset 
CFT are labelled by quadruples $(\Lambda,\lambda,m,\sigma)$, 
where $\Lambda$ stands for an integrable highest weight of ${\lieg}_k$, 
$\lambda$ for a weight of ${\lieh}$ and $m$ for the $U(1)$ charge. 
Furthermore, $\sigma$ is a weight of the $SO(2d)$ factor,
which is the vacuum, $0$, or the vector, $v$, in the NS sector,
and the spinor, $s$, or conjugate spinor, $c$, in the R sector.
The restrictions and identifications on the labels depend on
the particular coset one is considering. For our purposes, they
can be formally implemented by considering a simple current
extension \bert\ of the tensor product,
$$
\left[\lieg\times SO(2d)\times\lieh^*\times U(1)^*\right]_{\rm extended}\,.
\eqn\tenpro
$$ 
At least for the modular properties of the model, this extended
tensor product is equivalent to the original coset model.
Since only modular data and with them the fusion
rules enter the construction of Cardy boundary states \cardy, this is
sufficient for our purposes.

Let us make this procedure concrete for the cosets $SU(n+1)_k/SU(n)$,
as an example. The extension is by the simple current 
$${\cal J} \ =\
(J^{(n+1)},J^{(n)},h,v)
\eqn\simpc
$$
in the tensor product \tenpro. Here, $J^{(n\!+\!1)}$ (respectively
$J^{(n)}$) denotes the generator of the cyclic simple current group of
$SU(n\!+\!1)_k$, (respectively $SU(n)_k$). Its monodromy charge,
$Q_{J^{(n\!+\!1)}}(\Lambda)=\tau_{n\!+\!1} (\Lambda)/(n\!+\!1)$ measures
the $(n\!+\!1)$-ality of the representation $\Lambda$ (analogously
$\tau_{n}(\lambda)$ stands for the $n$-ality of the representation
$\lambda$). Moreover $J^{(n\!+\!1)}$ acts on $\Lambda$ to yield
$J^{(n\!+\!1)}\Lambda$, by rotating clockwise the Dynkin labels of the
corresponding highest weight of the affine Lie algebra $SU(n\!+\!1)_k$
(and similarly for $SU(n)$).
Extension by the simple current $\cal J$ is equivalent to the selection
rule
$$
Q_{(J^{(n+1)},J^{(n)},h,v)}(\Lambda,\lambda,m,\sigma) =
\frac{\tau_{n+1}(\Lambda)}{n+1} + \frac{\tau_n(\lambda)}{n} + 
\frac{m}{n(n+1)} + Q_v(\sigma) \, =0 \bmod\zet,
$$
where $Q_v(\sigma)$ is $0$ in the NS sector and
$1/2$ in the R sector, and to the order $n(n+1)$ identification
$(\Lambda,\lambda,m,\sigma)\equiv {\cal J}(\Lambda,\lambda,m,\sigma)
\equiv (J^{(n+1)}\Lambda,J^{(n)}\lambda,m+h,v\sigma)$.
We refer to the literature for further details.

It is well-known \LVW\ that the ring of primary chiral fields of any
one of these models \KSdef\ is isomorphic to the cohomology ring of the
underlying Grassmannian,
$$
\rnk n{k}\ \cong\ \Hst\big(\,\coeff{SU(n\!+\!k)}{SU(n)\times
SU(k)\times U(1)},\IR\,\big)\ ,
\eqn\ringiso
$$
with
$$
{\rm dim} (\rnk n{k})\ =\  \left({n+k \atop n}\right)\ .
\eqn\dimR
$$
The relations in this ring can be integrated to a potential  $\wnk
nk(x_i)$, which can be interpreted as superpotential of a \LG model
with fields $x_i$, $i=1,\dots, n$ (with $U(1)$ charges
$q_i=i/(\npko)$).  The superpotentials were explicitly given in
\doubref\LVW\DGfr\ and can be compactly characterized by the
following generating function:
$$
-\log\Big[\sum_{i=1}^{n-1}(-t)^ix_i\,\Big]\ =\
\sum_{k=-n+1}^{\infty}t^{n+k}\,\wnk nk(x_i)\ .
\eqn\genallW
$$

The quasi-homogenous superpotentials $\wnk nk(x_i)$ represent isolated
singularities that can be viewed as generalizations of the $A_{k+1}$
simple singularities; those were discussed in the introduction and
correspond to $\wnk 1k(x_i)$. In analogy to the minimal models and
their relationship to ALE spaces, we will be interested in comparing
the $D$-brane geometry of the resolved singularities:
$$
\wnk nk(x_i,\mu)\ =\ \wnk nk(x_i)+\mu\ 
\eqn\resolvedWnk
$$
to the boundary CFT of the coset models. The resolution is
distinguished in that it preserves the discrete $\ZZ_{h}$ ($h\equiv
\npko$) ``Coxeter'' symmetry that is intrinsic to the coset models.

The resolved potential \resolvedWnk\ can be viewed as the inhomogenous
form of a \LG potential for a non-compact Calabi-Yau space. However,
the most natural way to form such a space is {\it not} to tensor an
\nex2\ coset model with the \nex2 Liouville theory as in \JSgor, 
because this would
generically require fractional powers of LG fields. Rather, the most
natural way is to tensor the coset model with a matching, generalized
Liouville theory with $n$ fields $z_i$ (with charges $q_i=-i/h$). The
combined system has central charge
$$
\hat c(n,h)+\hat c(n,-h)\ =\ 2n\ , 
\eqn\centralC
$$
(where $\hat c(n,h)\equiv (h-n-1)n/h$), 
which corresponds to a non-compact $2n$-fold. 

In the next sections we will compute intersection indices
$\I_{a,b}\equiv {\rm Tr}_{a,b}[(-1)^F]$ between boundary states $a$,$b$
of the \nex2 coset models.  They will gain a concrete
geometrical meaning only after taking the non-compact piece into
account, which produces symmetric generalized Cartan matrices:
$$
\hat C\grl nK\ =\ \Ih\grl nk+\big(\Ih\grl nk\big)^t\ .
\eqn\symmCh
$$
However our main concern will be the intrinsic
properties of the boundary CFT of the \nex2 coset models.
(Note that we have put the hat to indicate that we will
obtain extended Cartan matrices associated with 
over-complete, $\ZZ_h$ symmetric homology bases.)

\chapter{Boundary states and their intersection index in \nex2 coset models}

Before we start computing the intersection index from the BCFT,
we would like to make a few comments on the general class
of boundary conditions we will consider. The conformal field theories
of our interest are rational in the closed string sector, with respect
to  an extended chiral algebra (given by \nex2\ $W$-algebras).
Because of lack of appropriate CFT methods as of now, we will have to
use rationality also in the open string sectors. This means that we
will not be able to specify all \nex2\ superconformal boundary
conditions, and thus won't get all boundary states. The boundary
conditions we will obtain are only those that preserve the
full chiral algebra of the models, and this is generically only a small
subset of all possible \nex2 supersymmetric ones.

To be precise, we will consider A-type (with respect to the  \nex2\
algebra) boundary conditions, using the charge conjugation  modular
invariant in the closed string sector. Our boundary conditions will
thus preserve the full chiral algebra without twist. From the CFT point
of view of the maximally extended chiral algebra, this is commonly
referred to as the ``Cardy'' case. The Cardy boundary states are
labelled in the same way as the the primary fields are, namely by
(orbits of) $(\Lambda,\lambda,m,\sigma)$ with the same selection and
identification rules.

We have outlined some general conformal field theoretic features of the
intersection index in Appendix \appA. We show there in particular that
it can be written in terms of the annulus coefficients, $A_{ab}^m$,
as follows:
$$
\I_{ab} = \sum_{m\; {\rm Rgs}} A_{ab}^{s^{-1}m} - A_{ab}^{vs^{-1}m} \,,
\eqn\inop
$$
where $v$ denotes the simple current corresponding to the worldsheet 
supercurrent and $s$ the simple current corresponding to spectral flow
by half a unit. The sum in \inop\ is over all Ramond ground states $m$. 
Thus, the $s^{-1}m$ are chiral primary fields. In the cases of our 
interest, \inop\ simplifies further since the annulus coefficients 
are simply identical to the fusion coefficients, i.e., the structure 
constants of the Verlinde algebra of the coset model. Modulo field 
identification fixed points, those are given by the products of fusion 
coefficients of the factors in \tenpro, restricted to allowed fields, 
and summed over field identification orbits.
We will denote the fusion coefficients of ${\lieg}$ and ${\lieh}$, 
by $\NG$ and $\NH$ respectively. The fusion coefficients 
of the $U(1)$ factor are conveniently encoded in a shift matrix
$g$. The fusion coefficients of the $SO(2d)$ factor are
defined by $vs=c$, $v^2=0$, $s^2=v^{d}$.

We view the intersection numbers of boundary states with
representatives $(\Lambda_1,\lambda_1,m_1,\sigma_1)$  and
$(\Lambda_2,\lambda_2,m_2,\sigma_2)$, for fixed $\Lambda_1$ and 
$\Lambda_2$, as a matrix in $\lambda_1,m_1$ and $\lambda_2,m_2$. 
Let us also fix $\sigma_1=\sigma_2=0$. From \inop\ we have
$$
\left(\Ih_{\Lambda_1,\Lambda_2}\right)^{\lambda_1,m_1}_{\lambda_2,m_2}
=\!\!\! \!\sum_{(\Lambda,\lambda,m,\sigma) {\;\rm ch.\ prim.}}  
\!\!\! \! 
\Ncos_{(\Lambda,\lambda,m,\sigma)(\Lambda_2,\lambda_2,m_2,0)}^{%
(\Lambda_1,\lambda_1,m_1,0)}
- \Ncos_{v(\Lambda,\lambda,m,\sigma)(\Lambda_2,\lambda_2,m_2,0)}^{%
(\Lambda_1,\lambda_1,m_1,0)},
$$
where the sum is over all chiral primary field representatives.
Insert the fusion coefficients of $\lieg$, $\lieh$,  $U(1)$,
and $SO(2d)$ then gives
$$
\eqalign{
\left(\Ih_{\Lambda_1,\Lambda_2}\right)^{\lambda_1,m_1}_{\lambda_2,m_2}
&= \sum_{\Lambda} \NG_{\Lambda\Lambda_2}^{\Lambda_1}\ \times\ \cr
&\!\!\left[
\sum_{\lambda,m\atop
(\Lambda,\lambda,m,0) {\;\rm ch.\ prim.}}
\NH_{\lambda\lambda_2}^{\lambda_1} (g^{-m})_{m_2}^{m_1}
-
\sum_{\lambda,m\atop
(\Lambda,\lambda,m,v) {\rm ch.\ prim.}}
\NH_{\lambda\lambda_2}^{\lambda_1} (g^{-m})_{m_2}^{m_1}
\right]
}\,,
\eqn\inzane
$$

We see that we need to know which $\lambda,m$ labels yields, for
fixed $\Lambda$, a chiral primary field. To this end, we use the fact 
that any Ramond ground state has a representative
$(\Lambda,\lambda,m,\sigma)$ with
$$
(\lambda,m) + (\rho_\lieh,0) = \weyl (\Lambda + \rho_\lieg)\ ,
\eqn\relweyl
$$
where $\rho_\lieh$ and $\rho_\lieg$ are the Weyl vectors, and where
$\weyl$ runs over the minimal length representatives, $W(G/H)$, of the 
Weyl group coset $\relweylgroup$. The $w\in W(G/H)$ can also be 
uniquely characterized by the fact that $\lambda$ is an integrable 
highest weight of $\lieh$ at the level of interest. In \relweyl, 
$m$ is determined by the embedding of the $U(1)$ factor
in $\lieg$, and the $SO(2d)$ representation $\sigma$ is the spinor, $s$, 
or conjugate spinor, $c$, if the sign of $\weyl$ is $+1$ or $-1$, 
respectively. Using spectral flow to the NS sector, given by
$(0,0,m_0,s)$, for a particular $m_0$, we see that a solution to
\relweyl\ contributes in \inzane\ with a sign equal to $\weylsign$.

However, not all Ramond ground states representatives are of the form 
\relweyl. We also have to implement the identification
rules that do not change a given $\Lambda$. They introduce an additional 
sign if they act non-trivially on the $SO(2d)$ label. Summing up,
we can write \inzane\ in the compact form
$$
\left(\Ih_{\Lambda_1,\Lambda_2}\right)^{\lambda_1,m_1}_{\lambda_2,m_2}
= \sum_{\Lambda} \NG_{\Lambda\Lambda_2}^{\Lambda_1}
\sum_{w\in W(G/H)}
\sum_{(\lambda,m)}{}^{'}
\epsilon \;
\weylsign
\NH_{\lambda\lambda_2}^{\lambda_1} (g^{-m+m_0})_{m_2}^{m_1}\ .
\eqn\compactI
$$
where $\sum'$ is over all those $(\lambda,m)$ that are related
to \relweyl\ by a field identification in the
denominator and in the $SO(2d)$ factor (which determines the
sign $\epsilon=\pm 1$).

\section{Examples}

As a first example, 
reconsider the intersection of the $\Lambda\equiv\ell=0$ states of
the \nex2\ minimal models, $SU(2)_k\times SO(2)_1
/U(1)_{2h}$. Here, $W(G/H)=W(SU(2))$ consists just of two elements,
namely of the identity $\weyl_0(l)=l$ and of $\weyl_1(l)=-l$.
Furthermore, $m_0=1$, and $\weyl_0(0+\rho_{SU(2)})-m_0 = 0$,
$\weyl_1(0+\rho_{SU(2)})-m_0 = -2$, so that there are two
terms in the intersection matrix:
$$
\Ih_{0,0}\grl1k = 1-g^{2}\ .
$$
This reproduces the result \Inullnullext\ (modulo reducing
the size of the matrix $g=\ga(2h)\equiv g_{(2(k+2))}$ in order
to avoid redundancy).

The second example we consider are the Kazama-Suzuki models
$SU(3)_k/U(2)$. The full coset model reads
$$
\frac{SU(3)_k\times SO(4)_1}{SU(2)_{k+1}\times U(1)_{6h}}\ ,
$$
where $h=k+3$ and the ``level'' of the $U(1)$ is related to the radius
of the compact boson in the usual way. Primary fields in the coset are
labelled by allowed field identification orbits of
$$
\left((l_1,l_2),\lambda,m,\sigma\right)\ ,
$$
where $l_1,l_2,\lambda\ge 0$, $l_1+l_2\le k$,
$\lambda\le k+1$, $m$ is defined modulo $6h$ and $\sigma$ is scalar
($0$) or vector ($v$) in the NS sector and spinor ($s$) or conjugate
spinor ($c$) in the R sector.

Let us fix $(l_1',l_2')$ and $(l_1'',l_2'')$,
and consider boundary states with varying $\lambda$ and $m$,
$\sigma=0$. Then the intersection matrix of those states is
$$
\Ih_{(l_1',l_2'),(l_1'',l_2'')}
= \sum_{(l_1,l_2)} {\cal N}_{(l_1,l_2)
\;\;(l_1'',l_2'')}^{\quad(l_1',l_2')} \;\;
\tilde \I_{(l_1,l_2)}\ ,
\eqn\finter
$$
where the ${\cal N}$'s are the $SU(3)_k$ fusion coefficients, and
$\tilde \I_{(l_1,l_2)}$ is the contribution
of all ground states in the open string R sector
that can occur for fixed $(l_1,l_2)$, modulo
field identification. This reads explicitly
$$
\eqalign{
\tilde \I_{(l_1,l_2)}
 & = N_{l_1} g^{-l_1-2l_2}
 - N_{l_1+l_2+1} g^{-l_1+l_2 +3} \cr
 &\qquad\quad+ N_{l_2} g^{2l_1+l_2 +6} 
 -N_{k+1-l_1} g^{-l_1-2l_2+3h}   \cr
 &\qquad\qquad\qquad + N_{k-l_1-l_2} g^{-l_1+l_2 +3 +3h}
 - N_{k+1-l_2} g^{2l_1+l_2 +6+3h} \cr
 &= 
 \bigl( N_{l_1} g^{-l_1-2l_2}
 - N_{l_1+l_2+1} g^{-l_1+l_2 +3}
 + N_{l_2} g^{2l_1+l_2 +6} \bigr)
(1-N_{k+1} g^{3h})
\ .
}
\eqn\inter
$$
Here and from now on, the $N$'s will be reserved to denote the 
$SU(2)$ fusion matrices. The matrix $g$ is the
$6h\times 6h$ dimensional basic shift matrix. The terms on 
the RHS of \inter\ correspond, respectively, to the occurrence 
of the fields
$$
\eqalign{
 &\bigl((l_1,l_2),l_1,l_1+2l_2,0\bigr) \cr
 &
 \bigl((l_1,l_2),l_1+l_2+1,
 l_1-l_2-3,v\bigr) \cr
 &\qquad\qquad\qquad\equiv
 \bigl((k-l_1-l_2,l_1),k-l_1-l_2,
 k+l_1-l_2,0\bigr) 
 \cr
 &\bigl((l_1,l_2),l_2,-2l_1-l_2-6,0\bigr) \cr
 &\qquad\qquad\qquad\equiv  
 \bigl((l_2,k-l_1-l_2),l_2,
 2k-2l_1-l_2,0\bigr) 
 \cr
 &\bigl((l_1,l_2),k+1-l_1,l_1+2l_2+3h,v\bigr) \cr
 &\qquad\qquad\qquad\equiv 
 \bigl((l_1,l_2),l_1,l_1+2l_2,0\bigr) 
 \cr
 &\bigl((l_1,l_2),k-l_1-l_2,
 l_1-l_2-3+3h,0\bigr) \cr
 &\qquad\qquad\qquad\equiv
 \bigl((k-l_1-l_2,l_1),k-l_1-l_2,
 k+l_1-l_2,0\bigr) 
 \cr
 &\bigl((l_1,l_2),k+1-l_2,-2l_1-l_2-6+3h,v\bigr)\cr
 &\qquad\qquad\qquad\equiv 
 \bigl((l_2,k-l_1-l_2),l_2,
 2k-2l_1-l_2,0\bigr) 
}
$$
in the open string sector. According to \inop, the fields with 
$\sigma=0$ contribute with a plus sign and the fields with $\sigma=v$
with a minus sign; this explains the signs in \inter. The structure
of \inter\ is as expected from \compactI. The first bracket is
the sum over the relative Weyl group, while the second implements
the identification trivial in the numerator of the coset.

\section{Properties of the intersection index}

We now analyze some of the properties of the
intersection index in Kazama-Suzuki models, as obtained
from the CFT computations. We will mainly work out the details
for the $SU(3)_k/U(2)$ models, but also indicate how they
generalize to more general coset models.

The Cardy construction provides us with a list of boundary  states
labelled by the primary fields of the Kazama-Suzuki model, and above we 
have computed the intersection index $\Ih$ between any pair of them. The
intersection index gives the set of boundary conditions the structure
of an integral lattice. As we will see in a moment, the rank of the
intersection form is given by the dimension \dimR\ of the chiral ring.
To reduce the size of the lattice, it is natural to look for an integer
basis amongst the states with $\Lambda=0$. Indeed we will find that all
other states can be obtained by integral linear combinations of (a subset
of) the $\Lambda=0$ states. These are thus the analogs of the basic
$\ell=0$ states of the minimal models, and in fact they correspond to
the $D$-brane states with lowest mass if we resolve the singularity by
switching on $\mu$ in \resolvedWnk.

From the formulae above, it is obvious that a state with 
(representative) label $(\Lambda,\lambda,m,0)$ intersects all 
other states with a minus sign relative to the state 
$(\Lambda,\lambda,m,v)$ (brane and anti-brane). Thus, we can
immediately restrict our attention to, say, $\sigma=0$ states. 
Furthermore, in many instances there are identification rules 
that are trivial in the numerator of the coset, and this leads to 
a further reduction of the labels among $\Lambda=0$
representatives.

Let us make this explicit for our favorite example, $SU(3)_k/U(2)$. 
From \finter\ and \inter, we deduce the basic intersection
matrix of the states with $\Lambda=0$ representatives:
$$
\eqalign{
\Ih\grl2k\ \equiv \Ih_{(0,0)(0,0)}\grl2k  &=  1- N_1 g^{3} + g^{6} 
 - N_{k+1}g^{3h} + N_{k} g^{3h+3} - N_{k+1} g^{3h+6} \cr
 &=  (1-N_{k+1} g^{3h}) (1-N_1 g^{3} + g^{6}) 
}
\eqn\basic
$$
Suppressing the $\Lambda=(0,0)$ label, the remaining labels are
$(\lambda,m,0$). Note that for the $\Lambda=0$ states, $m$ is always a
multiple of three, and we can therefore reduce the size of the
$g$-matrix accordingly: $g=g_{(6(k\!+\!3))}\to g_{(2(k\!+\!3))}$. The coset
rules require $\lambda/2+m/6$ to  be integer, and moreover identify
$(\lambda,m,0)$ with $(k+1-\lambda, m+3h,v)$. Therefore, we can
restrict ourselves to the following ``standard'' range:
$$
\lambda=0,...,k+1, m = 3 m' \;{\rm with}\; m'= \lambda,..
2k+2 - \lambda \;{\rm and}\; \lambda-m' \;{\rm even}.
$$
If we write
$$
\eqalign{
l_1' &= \lambda \cr
l_2' &= \frac{\lambda - m'}{2}\ ,
}
$$
the standard range can be more concisely expressed as:
$$
l_1',l_2' \ge 0\ ,\qquad l_1'+l_2'\le k+1\ .
\eqn\standardR
$$ 
This formally looks like the labels of 
the integrable representations of $SU(3)_{k+1}$ 
(where the level is by one higher than what appears in the coset),
however we will later see that the labels should be
interpreted in terms of the representations of $U(2)$.

It is easy to see that restricting the labels to $l_1'+l_2'\le k$,
which corresponds to the integrable representations of $SU(3)_{k}$, and
ordering the states according to increasing $l_2'$ and $l_1'$,  the
reduced intersection form is upper triangular with $1$ on the diagonal;
We denote it by omitting the hat: $\I\grl2k\equiv 
\I_{(0,0)(0,0)}\grl2k$. It has rank equal to
$(k+1)(k+2)/2$, which is equal to the dimension of the chiral
ring of $SU(3)_k/U(2)$.

The $\Lambda=0$ boundary states with $l_1'+l_2'\le k$ thus yield a
complete basis of the charge lattice, and what remains to be shown is
that all other boundary states can be obtained from them via integral
linear combinations. As far as the rest of the $\Lambda=0$ states is
concerned, namely the ones with $l_1'+l_2'= k+1$, this can be seen in
the following way. Simply observe that the sums of states
$$
(0,l_2') + (1,l_2') + ... + (k+1,l_2') 
$$
(assuming they are mapped backed to the standard range with an
appropriate minus sign) do not intersect with any other state, and so
correspond to null eigenvectors of $\Ih$. This shows in a direct
way that the states with $l_1'+l_2'= k+1$ can be written as integral linear
combinations of the states with $l_1'+l_2'\le k$. As for the remaining
states with  $\Lambda> 0$, we show in Appendix \appB\ that also the
charges of these states can be expressed as integral linear
combinations of the $\Lambda=0$ states.

The above considerations can be made more transparent by associating a
graph with the basic intersection index  \basic, whose nodes correspond
to boundary states and oriented signed links between them encode their
intersection. We have displayed the graph (omitting the arrows) 
for $k=2$ in \lfig\graphst. In this picture, 
the fat lines denote the sub-graph $\I\grl 22$  of
the integral homology basis, which corresponds to the fusion graph of
$SU(3)_2$ (by change of basis it can be put into the form of the $D_6$
Dynkin diagram, which reflects the equivalence of the KS model
$SU(3)_2/U(2)$ with the minimal model of type $D_6$).
Note that the  extended graph looks similar to the
fusion graph of the integrable representations of $SU(3)_3$, but we
will see later that the dashed links really make it into a fusion
graph of $U(2)$.

\figinsert\graphst{The intersection graph $\Ih\grl 22$  of
$\Lambda=0$ boundary states of  the $SU(3)_2/U(2)$ KS model. The fat
lines denote the sub-graph $\I\grl 22$  of the integral
homology basis, which coincides with the fusion graph of
$SU(3)_2$. The open dots denote extending nodes (in
analogy to $\gamma_0$ in \lfig\AnDynkin), which give the fusion graph
of $SU(3)_3$; as we will see later, 
the dashed links extend this further to the
fusion graph of $U(2)$.}{1.5in}{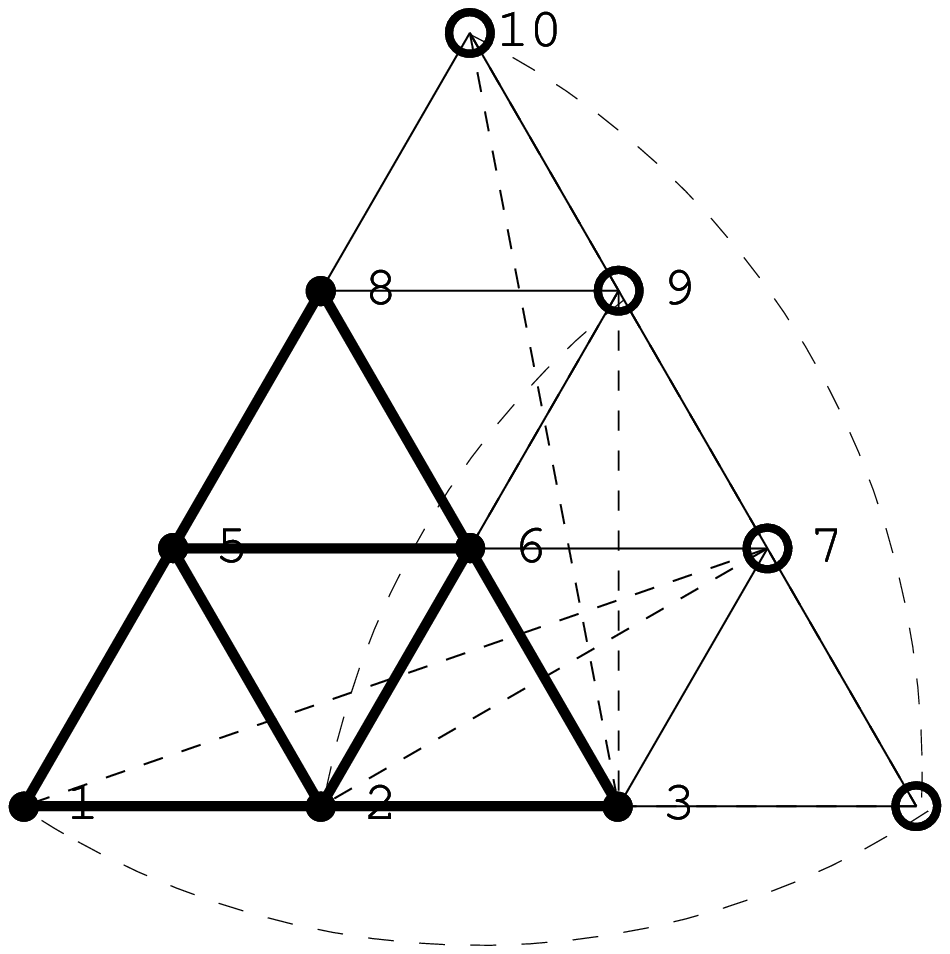}

The generalization of \basic\ to all KS models of the form 
$SU(n+1)_k/U(n)$ is straightforward. The $\Lambda=0$, $\sigma=0$ states 
intersect as
$$
\eqalign{
\Ih\grl nk \equiv \Ih_{\vec0,\vec0}\grl nk &=\ 
1 - N_{[1]} g^{n+1} + N_{[2]} g^{2(n+1)} + .... + (-1)^{n} g^{n(n+1)} \cr
&+
(-1)^{n+1}N_{J} g^{-(n+1)h} + (-1)^{n+2} N_{J[1]} g^{-(n+1)h+(n+1)} 
+ \dots 
\cr &\qquad\qquad\qquad\qquad\ \ \ +(-1)^{2n+1} N_{J} g^{-(n+1)h+n(n+1)} \cr
\vdots& \cr
&+
(-1)^{(n+1)(n-1)} N_{J^{n-1}} g^{-(n+1)(n-1)h} + \dots 
\cr &\qquad\qquad\qquad\qquad\ \ \ + (-1)^{(n+1)(n-1)+n}
N_{J^{n-1}} g^{-(n+1)(n-1)h +n(n+1)}
\cr
&= \left(1 - N_{[1]} g^{n+1} + N_{[2]} g^{2(n+1)} + .... + (-1)^{n} 
g^{n(n+1)}\right)\cr
&\times \left(1+(-1)^{n+1}N_{J}g^{-(n+1)h} + 
\dots + (-1)^{(n+1)}N_{J^{n-1}} g^{-(n+1)(n-1)h}
\right)
}
\eqn\general
$$
Here, $N_{[i]}$ is the fusion matrix of the $i$-th fundamental
representation of $SU(n)$ at level $k=h-n$, and $(0,J^{(n)},
(n+1)h,v^{n+1})={\cal J}^{n+1}$ is the simple current implementing the coset
rules that act only in the denominator, with $N_{J}$ the fusion matrix 
of $J\equiv J^{(n)}$.  Due to redundancy, the $U(1)$ fusion matrix
$g\equiv g_{(n(n+1)h)}$ can be reduced in size by a factor of $n\!+\!1$.

Similarly to the $SU(3)$ example discussed above, the coset
identification rules allow the reduction of the $\Lambda=0$ states to a
set of labels in one-to-one correspondence with the integrable
representations of $SU(n+1)_{k+1}$, which is at one level higher than
the CFT suggests.  The intersection matrix $\Ih\grl
nk$ does not have full rank and thus should be viewed as an
intersection form of an over-complete basis. Restricting to boundary
states corresponding to level $k$, the resulting reduced
intersection matrix intersection matrix $\I\grl nk$
becomes upper triangular and has full rank (given by
\dimR). The vanishing relations are
analogous to the $SU(3)$ case, and we have found a basis for the
charge lattice also in the general case.

Note that the graph of the symmetrized reduced matrix 
$\I\grl nk$,
$$
C\grl nk\ =\ \I\grl nk+\big(\I\grl nk\big)^t\ ,
\eqn\symmC
$$
which represents the intersection index for a complete homology basis,
coincides with  the fusion graph of $SU(n+1)_k$; this generalizes the
coincidence of the $A_{n+1}$ Dynkin diagram with the $SU(2)_k$ fusion
diagram as discussed in the introduction.
This also reproduces and clarifies, from a BCFT point of view, the
connection between the resolution of the singularities $\resolvedWnk$
and the Verlinde fusion  algebra for $SU(n+1)_k$. 
This relation had been conjectured by Zuber \JZg\ and 
others and was proven in \GZAV.

However, note that it is the extended intersection form
$\Ih\grl nk$ that is adapted to the $\ZZ_h$ Coxeter
symmetry of the coset models. In fact, as we will show below, it is the
more interesting and natural object to study, leading to a
generalization of the $\ZZ_h$ symmetric McKay quiver.

\section{Boundary fusion rings, quantum cohomology and soliton polytopes}

In the previous section we have obtained the general formula \general\
for the extended intersection index  $\Ih\grl n{k}\equiv\Ih_{\vec 0,\vec 0}\grl
n{k}$ of the basic $\Lambda=0$ boundary states of the KS models based
on $SU(n+1)_k/U(n)$. Recall that it can be written roughly as
alternating sum $\sum(-)^p N_p g^{pn}$, where $N_p$ are $SU(n)$ fusion
coefficients (or some simple current transforms thereof), and $g\equiv
g_{(nh)}$ is the reduced shift matrix \shift\ ($h\equiv\npko$). 
In fact one can rewrite it in the form
$$
(\Ih\grl n{k})_{a}^{\ b}
\ \equiv\
(\Ih_{\vec 0,\vec 0}\grl n{k})_{a}^{\ b}
\ =\ 
\sum_{p=0}^n(-1)^p (\hat\Fp\grl n{k}_p)_{a}^{\ b}
\ =\
\langle \bar \rho_b|\sum_{p=0}^n (-)^p\varphi[p]|\rho_a\rangle
\ ,
\eqn\Inullnullgen
$$ 
which is analogous to what we showed in the introduction for the
$A_{n+1}$ minimal models. The sum runs over massless intermediate open
string states $\varphi[p]$ that are labelled by the antisymmetric
fundamental representations\foot{ These are one-to-one to the
generators of the chiral ring. For the intersection indices with
$(\Lambda,\Lambda')\not=(\vec 0,\vec 0)$, more general representations
contribute as intermediate states.} $[p]$ of $U(n)$; these are given by
Young tableaux of one column with at most $n$ boxes. The boundary
states are labelled by representations $\rho_a$ of $U(n)_{k+1,h}$ that
are denoted by Young tableaux with at most $k$ columns whose height is
at most $n$ boxes. The very same Young tableaux are known \hill\ to
denote bundles on Grassmannians, a connection that we will discuss
elsewhere. While they are also formally one-to-one to the
integrable representations of $SU(n)_{k+1}$,  the tensor product
coefficients we encounter:
$$
[p]\otimes \rho_a\ =\ \bigoplus_b (\hat\Fp\grl nk_p)_{ab}\,\rho_b\ ,
\eqn\Fprulegen
$$
are the fusion coefficients of $U(n)_{k+1,h}$ -- up to a subtlety
that we now explain.

The Verlinde fusion ring of $U(n)_{k+1,h}$ is given by the naive
representation ring of $(SU(n)_{k+1}\times U(1)_{nh})/\ZZ_n$, modulo an
ideal $I$. In order to be explicit, 
let us consider the fusion ring of
$U(2)_{k+1,h}$, which can be thought of as a quotient of the
representation ring of   $(SU(2)_{k+1}\times U(1)_{2h})/\ZZ_2$, as
has been discussed in detail in ref.\ \EWgrass. The latter is spanned
by the symmetric powers $V_m=(V_1)^{\otimes m}$ of the two dimensional 
fundamental representation of $SU(2)$, and by powers of the one-dimensional
representation $W$ of $U(1)$. The individual fusion rings are generated
by imposing $V_k=0$ and $W^{2h}=1$.  Clearly, in order to form $U(2)$
representations, we have to restrict to operators $V_mW^l$ with
$l+m=$even. Moreover, in order to obtain the fusion ring of $U(2)$ we
need to impose the extra relation $I:\ V_mW^l=V_{k+1-m}W^{l+h}$.

In our context, the quotienting by $I$
corresponds  to field identification in the coset model, which is
implemented by the simple current ${\cal J}^{n+1}$ \simpc. It acts
also on the $s$-labels (which denote the $NS$- and $R$-sectors), and in
particular involves a shift of $\Delta s=2n-2$. This implies that for
odd $n$ the field identification maps between branes, while for even $n$
it maps branes to anti-branes. As a consequence, we get an extra sign in
the expression for the ideal, so what we get from the CFT is the
deformed relation $I\!:\ V_mW^l=-V_{k+1-m}W^{l+h}$.
In effect, for even $n$ the tensor
product coefficients $\hat\Fp\grl nk_p$ are equal to the fusion
matrices of $U(n)_{k+1,h}$ only up to a sign flip of the
lower-triangular part.\foot{We can flip signs of appropriate boundary
states to the effect that all entries in $\Ih$ are zero or positive.
This amounts to dropping $(-1)^p$ in \Inullnullgen\ and  undoing the
sign flip of the lower-triangular entries in $\hat\Fp\grl nk_p$, which
turns them into the true fusion matrices of $U(n)$. This basis is equally
allowed and this shows that the sign flip is not really
important.}

In order to find explicit expressions for the fusion matrices
$\hat\Fp\grl nk_p$, we can make use of the
fact \EWgrass\ that the fusion ring of $U(n)_{k+1,h}$ is isomorphic to
the {\it quantum} cohomology ring of $\gkn n\npko$. As far as the
mathematical structure of this ring is concerned, it is known
\refs{\EWqh,\KI,\SCCVss,\CVqr}\  that it is isomorphic to the
cohomology ring $\rnk n{k+1}$ up to a certain deformation. It can be
most concisely expressed by the following perturbed superpotential:
$$
\hat\wnk nk(x_i,\mu)\ =\ \wnk n{k+1}(x_i)+\mu x_1\ .
\eqn\wtildedef
$$
The $U(n)_{k+1,h}$ fusion matrices are given by the structure constants
of the associated chiral ring at  the special point $\mu=(-1)^n$
\EWgrass. More specifically we just said that there should be a sign
flip for even $n$, so the correct statement is to say that the
matrices $\hat\Fp\grl nk_p$ in \Fprulegen\ are equal to the perturbed
chiral ring structure constants associated with \wtildedef, at the
special point $\mu=-1\ \forall n$.

There exists a simple canonical
construction of  the structure constants of the cohomology ring of
$\gkn n\npko$ (which we explain in Appendix \appC), and this makes it
easy to write down explicit expressions for the $\hat\Fp\grl nk_p$, 
and in turn for the intersection index.  
For example, consider the coset model based on
$SU(3)_2/U(2)$. According to what we just said, the fusion matrices
$\hat\Fp\grl 22_p$ $p=1,2$ are given by the perturbed ring structure
constants of the model $SU(3)_3/U(2)$;  we listed these explicitly in
Appendix \appC. We thus can immediately write down the extended
intersection matrix \Inullnullgen:
$$
\Ih\grl22\ =\ 1 -\hat\Fp_1\grl22+\hat\Fp_2\grl22
=
\pmatrix{ \aa(1) & \aa(-1) & \aa(0) & \aa(0) & \aa(1) & \aa(0) & \aa(0)
& \aa(0) & \aa(0) & \aa(    0) \cr \aa(0) & \aa(1) & \aa(-1) & \aa(0) &
\aa(-1) & \aa(1) & \aa(0) & \aa(0) & \aa(    0) & \aa(0) \cr \aa(0) &
\aa(0) & \aa(1) & \aa(-1) & \aa(0) & \aa(-1) & \aa(1) & \aa(    0) &
\aa(0) & \aa(0) \cr \aa(-1) & \aa(0) & \aa(0) & \aa(1) & \aa(0) &
\aa(0) & \aa(    -1) & \aa(0) & \aa(0) & \aa(0) \cr \aa(0) & \aa(0) &
\aa(0) & \aa(0) & \aa(1) & \aa(    -1) & \aa(0) & \aa(1) & \aa(0) &
\aa(0) \cr \aa(0) & \aa(0) & \aa(0) & \aa(0) & \aa(0) & \aa(    1) &
\aa(-1) & \aa(-1) & \aa(1) & \aa(0) \cr \aa(1) & \aa(-1) & \aa(0) &
\aa(0) & \aa(    0) & \aa(0) & \aa(1) & \aa(0) & \aa(-1) & \aa(0) \cr
\aa(0) & \aa(0) & \aa(0) & \aa(0) & \aa(    0) & \aa(0) & \aa(0) &
\aa(1) & \aa(-1) & \aa(1) \cr \aa(0) & \aa(1) & \aa(-1) & \aa(    0) &
\aa(0) & \aa(0) & \aa(0) & \aa(0) & \aa(1) & \aa(-1) \cr \aa(0) &
\aa(0) & \aa(    1) & \aa(-1) & \aa(0) & \aa(0) & \aa(0) & \aa(0) &
\aa(0) & \aa(1) \cr  }
\ ,\eqn\Itwotwo
$$
where the basis corresponds to the $U(2)$ reps $\{\cdot,\tableau{1}\sim
V_1W,\tableau{2}\sim V_2W^2,\tableau{3}\sim V_3W^3,\tableau{1 1}\sim
W^2,\tableau{2 1}\sim V_1W^3, \tableau{3 1}\sim \sim V_2W^4,\tableau{2
2}\sim W^4,\tableau{3 2}\sim V_1W^5,\tableau{3 3}\sim W^6\}$. The
intersection matrix has rank equal to six, which is the dimension of
the chiral ring $\rnk22$. The graph of this intersection form is as
given in \lfig\graphst, but can also be represented in a $\ZZ_5$
symmetric form as in \lfig\graphsf\ (in this figure we chose a
different basis by flipping the signs of the 3th, 4th and 7th boundary
states, in order to have a manifest $\ZZ_5$ symmetry).


\figinsert\graphsf{Graph of the boundary CFT intersection form
$\Ih_{0,0}\grl 22$, equivalent to the $U(2)$ fusion graph of
\lfig\graphst. It is represented here in a manifestly $\ZZ_5$ symmetric
fashion, for which the links correspond to fundamental
solitons of the LG potential
$\hat\wnk 22(x,\mu)={1\over6}{x_1}^6 - {{x_1}^4}\,x_2 + {3\over2}
{{x_1}^2}\,{{x_2}^2} - {1\over3}{{x_2}^3} +\mu x_1$. The solid lines
denote intersections equal to $-1$ and the dashed lines intersections
equal to $+1$, while the open dots are the extending nodes. The fat
lines denote the fusions generated by $\hat\Fp\grl 22_1$ and the thin
lines those of $\hat\Fp\grl 22_2$.}{1.8in}{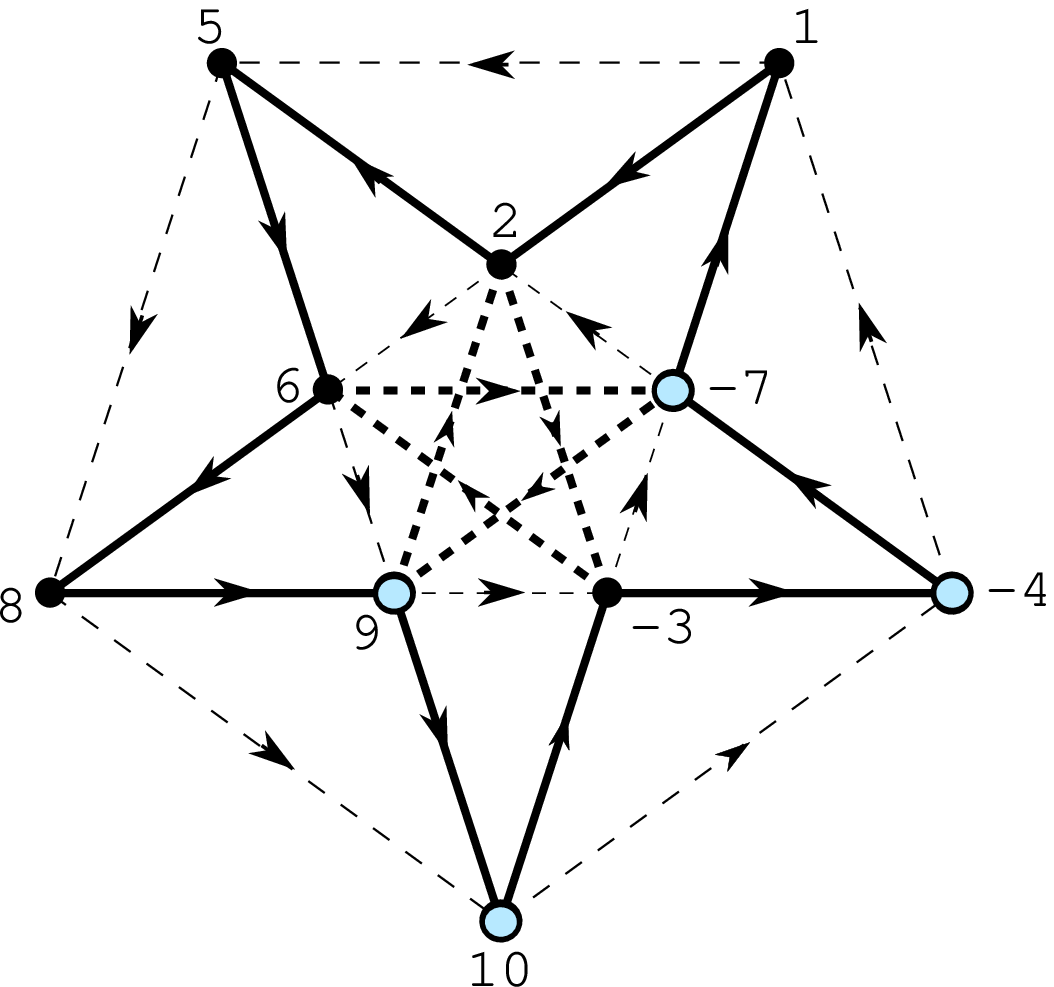}

It is easy to see that the links generated by $\hat\Fp\grl 22_1$
(fat lines in \lfig\graphsf) and by $\hat\Fp\grl 22_2$ (thin lines in
\lfig\graphsf) correspond to the fusions with the fundamental
representations $\tableau{1}\sim V_1 W$ and $\tableau{1 1}\sim W^2$,
respectively. Obviously  the fusions generated by $\hat\Fp\grl 22_2$
act cyclically, and thus realize the representation ring of
$\ZZ_5$. This is in analogy to what we discussed in section 1 for the ALE
space, and the question arises whether we can find here a
generalization of the McKay correspondence. Indeed \Inullnullgen\ and
\Fprulegen\ relate the intersection homology of the resolved singularity 
$W\grl nk-\mu=0$ \resolvedWnk\ to the fusion ring of $U(n)_{\npk,h}$,
and what remains to check is whether this fusion ring can
be expressed as representation ring of some discrete group, $\Gamma$.

However, while this is true for the fusions generated by $\hat\Fp\grl
22_2$, it is not true for the fusions generated by $\hat\Fp\grl 22_1$
(again, the fat lines in \lfig\graphsf). An easy way to see this is to
note that in tensor products the dimensions of the representations must
add up correctly.  Due to the $\ZZ_5$ symmetry of the diagram,  we have
a priori as free parameters the dimensions $d_i$ of the  nodes of the
inner circle and the dimensions $d_o$ of the outer circle, plus the
dimension $d_{\tableau{1}}$ of the fusing representation. From the
$\ZZ_5$ symmetry it suffices to consider the action of  $\hat\Fp\grl
22_1$ on one outer node and on one inner node. This leads to the
equations:
$$
\eqalign{
d_{\tableau{1}}\cdot d_o\ &=\ d_i \cr
d_{\tableau{1}}\cdot d_i\ &=\ d_o+d_i\ , \cr
}\eqn\dimrule
$$
which do not have an integer solution. This implies that the fusion
ring, which by definition is a truncated representation ring of $U(n)$,
cannot  be written for $n>1$ as a representation ring of some discrete group
$\Gamma$, and our search for a naive generalization of the McKay
correspondence for $n=1$ to arbitrary $n$ did not succeed -- perhaps not
unexpectedly, because the known generalizations of the McKay
correspondence \ITNA\ deal with orbifold singularities obtained by
modding out discrete groups $\Gamma$, while in contrast the
singularities  \resolvedWnk\ whose resolution we study here are in
general no orbifold singularities.\foot{An intrinsic relation between
\nex2 coset models and discrete groups that has been long sought for
(see e.g., \refs{\JZg,\JSmc,\JSgor,\zub} is thus still elusive; this
problem was one of the motivations for our study.}

On the other hand, precisely in line of what we discussed for the  ALE
spaces, the $\ZZ_h$ symmetric diagrams are structurally analogous to
soliton graphs associated with the perturbed potentials \wtildedef\ of
two dimensional \nex2\ \LG models. The perturbation by the field of the
lowest charge is distinguished in that the corresponding massive LG
theories are integrable, and these have been thoroughly investigated in
this context \refs{\FMVW,\FLMW,\SCCVcn,\KI,\LW,\FI}. In particular in
\LW\ such soliton diagrams in the $\hat W$-plane (=central charge
plane)  were analyzed in some detail, where it was argued that the
irreducible solitons correspond to root vectors mapping between the
weights of the representation $\Xi$, as defined in Appendix \appC.
Moreover it was found that the $\hat W$-plane diagram itself is nothing
but a projection of the ``soliton polytope'' (consisting of the weights
of $\Xi$ and the roots linking them) to a particular eigenspace of the
Coxeter element of the Weyl group (namely one on which it acts as
$e^{2\pi i/h}$). The presently discussed boundary intersection graphs
of $\Ih_{\vec 0,\vec 0}$ are sub-diagrams, in which only those
``solitons'' appear that correspond to roots of grades $1,\dots,n$ mod
$h$, and not to all the roots. This is reflected in \Inullnullgen\
where $p$ runs only over the fundamental representations with up to $n$
boxes.

Note, though, that our graphs are not supposed to describe central
charges of $2d$ solitons, rather they encode properties of the
interactions on the boundary. The links describe maps induced by the
$U(n)$ fusion matrices $\hat\Fp\grl n{k+1}_{\tableau {1}}, \hat\Fp\grl
n{k+1}_{\tableau {1 1}},\dots$, and physically correspond to acting with
fermionic open string zero modes on the boundary states which yields
different boundary states.  These matrices thus represent a ``boundary
ring'', which happens to coincide with the chiral ring associated with the 
superpotential \wtildedef. The latter
has a close relationship to the resolved potential \resolvedWnk, whose
intersection homology we probe with the boundary CFT. Indeed, in
direct generalization of the minimal models \derivrel\  (where
$n\!=\!1$), the potential \resolvedWnk\
is given by a derivative \WLwgrav\ as follows:
$$
\eqalign{
\wnk nk(x_i,\mu)\ &=\
\coeff1{n+k}\,\nabla_x\,\hat\wnk n{k+1}(x_i,\mu)\ ,\cr
\nabla_x \ &\equiv \ \sum_{i=1}^{n-1}(n-i)\,x_{i-1}{\del\over\del
{x_i}}\ .
}\eqn\relallW
$$
This makes again contact to the findings of ref.\
\nickBLG\ where boundary superpotentials were introduced whose orders
are one degree higher as compared to the bulk superpotentials.

\chapter{Quiver representations}

By definition the intersection graphs we discussed so far represent
quiver diagrams of a category, consisting of BPS boundary states as
objects and of the massless fermionic open strings mapping between
pairs of them (for a concise introduction in the present context, see
e.g., the appendix of ref.~\DFR\ and also \yhe);  the $U(n)$ boundary
fusion rings correspond to path algebras on the quivers. 
While so far quivers have been mainly used in physics to encode properties of
the world-volume gauge theories of $D$-brane probes \MDGM, we do not
have in the present context the extra data needed (eg., structure of
$D$-terms) for such an interpretation;  these data can be specified
only when we embed the coset CFT's in a geometrical context (``Gepner
models''), but this is outside of the present scope. 
Rather we are interested here only in properties that
are intrinsic to the boundary coset models.

While the representation theory of generic quivers is
arbitrarily complicated, we do have some useful extra information that
we can extract from the boundary CFT, in particular the charges
of the Cardy boundary states (rather, expansion coefficients
with respect to the basic $\Lambda=0$ boundary states). 
We have listed closed formulae for the
charges of the boundary states of the $SU(3)_k/U(2)$ models in Appendix
\appB, and we will discuss some of their properties momentarily; it is
clear that our considerations generalize straightforwardly to other
cosets.

Before we discuss some features of the charges  in relation to the
quiver diagram in \lfig\graphsf, let us first digress and consider the charges
when projected to the minimal homology basis corresponding to the
reduced Cartan matrix $C$ in \symmC, with
$\I\grl22=1-\Fp_1\grl22+\Fp_2\grl22$. We have already mentioned that
upon change of basis, $C$ can be transformed to the Cartan matrix of
$D_6$. Scanning through the list of the $6\cdot10$ charge vectors that
we get from our formulae,  we find only 20 different projected charge
vectors (plus their negatives) as a consequence of field
identifications. As expected, these belong either to the positive 
or to the negative roots of $D_6$. 
While this is reassuring, we still miss ten more in
order to complete the set of positive roots. 
This reflects the limitations of the BCFT methods that
we employ for the analysis of boundary states in cosets. We have
only constructed the Cardy boundary states that are associated with
fully symmetric (wrt.\ the chiral \nex2\ $W$-algebra) boundary 
conditions. The ten missing states correspond to symmetry breaking
boundaries. For the special case $SU(3)_2/U(2)$, the symmetry
breaking boundary conditions can be constructed using methods of \FSC.
This happens because the model can be written as the
simple current extension of an \nex2\ minimal model
$SU(2)_8/U(1)$, which is still rational.\foot{It was shown in 
ref.~\LLS\ using slightly different methods how to obtain all 
boundary conditions, symmetry breaking and symmetry preserving, 
in this model, and more generally for $D$-branes on ALE
spaces of arbitrary ADE type.}

However, the situation becomes much worse for cosets at higher levels, which
generically have indefinite unextended Cartan matrices $C$. For example, for
$SU(3)_3/U(2)$ the ten dimensional unextended Cartan matrix  has two
zero eigenvalues and the charge lattice we get from the BCFT is of type
$E_8\times U\times U$, where $U$ corresponds to a null direction. This
is exactly as expected from the geometry of the triangle singularity
\bries\ of type $T_{2,3,6}$, to which the LG potential corresponds. The
degenerate intersection form should lead to a hyperbolic algebra with
infinitely many positive roots, however from the rational boundary CFT
we find only finitely many states 
(they turn out to be a subset of the
roots of $E_8$ plus a few imaginary null roots).

We now return considering the extended Cartan matrix 
$\hat C=\Ih\grl22+(\Ih\grl22)^t$ associated with the quiver in
\lfig\graphsf. The expected dimension of the moduli space of a 
quiver representation labelled by the charge vector $q$ is:
$$
d(q)\ =\ 1- \shalf q\cdot \hat C\cdot q\ .
\eqn\expdim
$$
We find that all boundary states we get in the $k=2$ model have $d=0$ and thus
correspond to rigid, indecomposable Schur roots of the quiver.
The $\Lambda=0$ states obviously correspond one-to-one to the ten nodes of
the quiver, while the states with $\Lambda\not=0$ correspond to certain
collections of nodes. It is reassuring to find that such collections are
always connected by at least one $-1$ link, like the one shown in
\lfig\boundst. This means that these states are good bound states of the
basic $\Lambda=0$ boundary states and not multi-brane states, exactly
as expected. We view this consistency as another successful
test of the consistency of the BCFT methods, applied here to a slightly
less standard situation.

\figinsert\boundst{
On the left part of the quiver diagram we depicted a BCFT state
belonging to the $\Lambda=\Lambda_1$ orbit which is a bound
state of the indicated basic states. On the right we show the
null state $\delta_1$ which has zero self-intersection and may be
viewed as a generalized $D0$-brane.
}{1.5in}{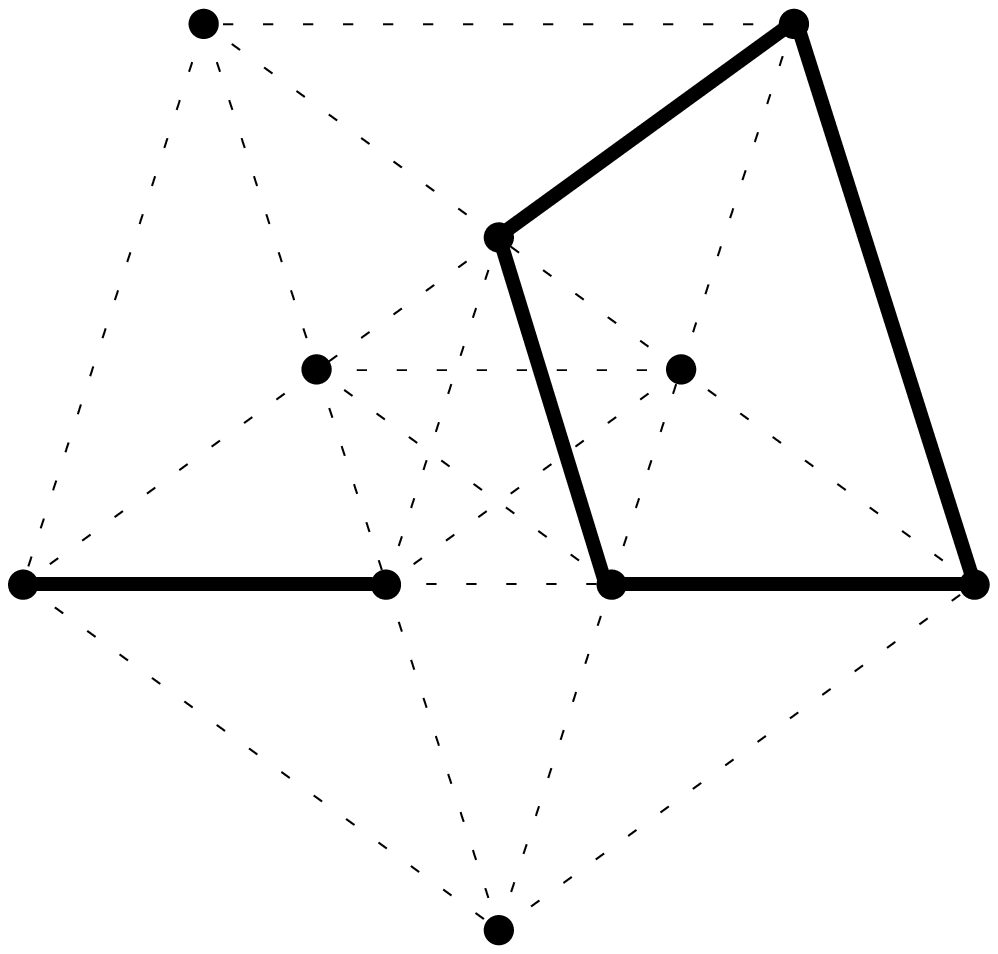}

An important feature of the extended Cartan matrix $\hat C$ is that it
has several null eigenvectors. In the quiver diagram these can be
associated with the nodes $(1,2,3,4)$, $(5,6,7,1)$, $(8,9,2,5)$,
$(10,3,6,8)$ and $(4,7,9,10)$, respectively; see again Fig.4 for an
example. We denote these null vectors, of which only four are linearly
independent, by $\delta_i$, $i=1\dots5$. The null vectors can be added to
any charge vector at no cost, ie., without changing the inner product
in \expdim.

This is analogous to the McKay quiver for the $A_{k+1}$ ALE space
shown in \lfig\AnDynkin, where adding the highest root 
$\delta=(1,1,..1,1)$ to the
charges does not change the dimension $d$. In physical terms, adding
$n\,\delta$ corresponds to bound states of $D2$-branes with $n$
$D0$-branes, and amounts to extending the set
of integrable representations to all representations of $SU(2)_k$; this
is depicted in the upper part of \lfig\affineD. 
This extension is well-known \doubref\naka\FM\
and one can indeed reproduce it from the BCFT fusion matrices as
pointed out in \LLS. However this is not to say that we have a
well-defined construction of the $D0$-brane states in the coset BCFT, at
least as far is known to us.

\figinsert\affineD{
On the top we see the $\hat A_{k+1}$ quiver spectrum for $k=3$. Each of
the leftmost three points corresponds to a $\ZZ_{k+2}$ orbit of boundary
states of the $SU(2)_3/U(1)$ coset model. The $n$-th repetition of this 
spectrum (not seen in the minimal CFT) 
corresponds to bound states with $n$ $D0$ branes. 
On the bottom we see the analogous picture for the
$SU(3)_2/U(2)$ theory. Each of the six leftmost points is associated to
an integrable representation of $SU(3)_2$, and labels an orbit of $10$
boundary states (modulo identifications). The other copies correspond
to generalized $D0$ brane bound states, the whole picture representing
the set of the affine highest weights of $SU(3)_2$.
}{1.9in}{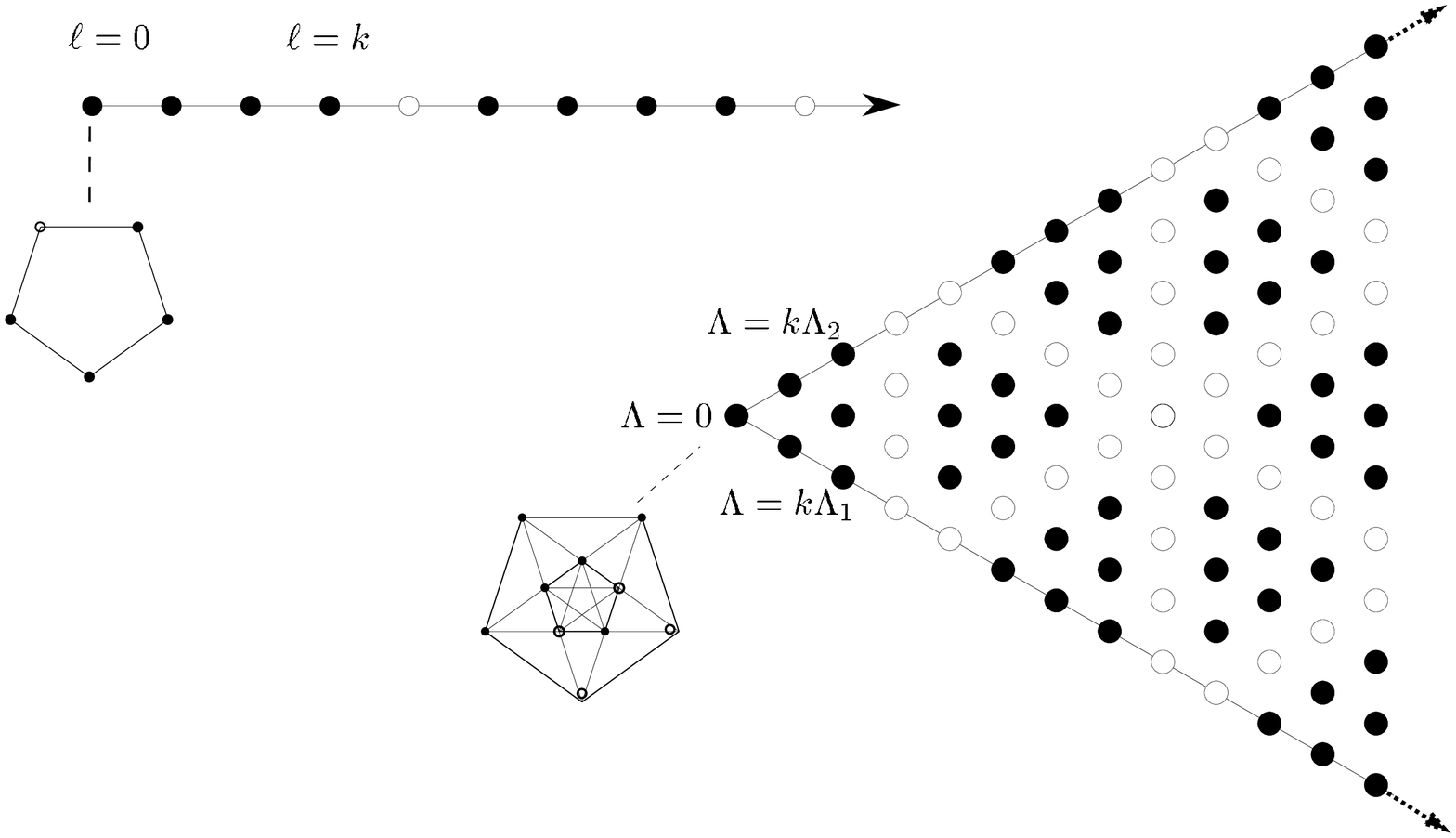}

In the presently discussed theories we find a generalization of this
structure. More precisely, let us focus on the formula (\appB.2) for the
boundary state charges and consider $SU(3)$ labels $\Lambda$ beyond 
the set of integrable representations at level $k$,
$R_{SU(3)_k}=\{\Lambda=\ell_1\Lambda_1+\ell_2\Lambda_2,\ \ell_i\geq0,\
\ell_1+\ell_2\leq k\}$. From the periodicity of the fusion
coefficients we then find that the charges are given by the
charges in the standard range $R_{SU(3)_k}$ 
shifted by linear combinations of the null vectors:
$$
Q_\Lambda\ =\ Q_\Lambda'+\sum n_i\,\delta_i\ ,\qquad\qquad \Lambda'\in
R_{SU(3)_k}\ .
\eqn\deltashift
$$
As the null vectors do not change the intersection properties,
we get an infinite repetition of the rational BCFT spectrum.
The whole structure we get is the affine Weyl alcove of 
$SU(3)_k$, which consists of all affine highest weights of 
$SU(3)_k$ and not of only the integrable ones. It can be characterized
by
$$
\hat R_{SU(3)_k}\ =\ \Big\{\Lambda=\ell_1\Lambda_1+\ell_2\Lambda_2,\
\ell_i\geq0,\ 
(\Lambda+\rho)\cdot \a_i\, \not=\, 0\, {\rm mod}\,
(k\!+\!3)\,\Big\}\ ,
\eqn\fullringLG
$$
where $\rho$ and $\a_i$ denote Weyl vector and roots of $SU(3)$. Note
that basically all affine weights $\Lambda$ are allowed, except that
there are planes of codimension one which correspond to orbits of null
states with zero self-intersections.

Precisely this structure has been found in the past in
refs.~\doubref\WLwgrav\topw\ in the context of coupling \nex2 coset
models to  topological gravity and related integrable systems; there
the infinite extension corresponds to the extension of the bulk chiral
ring by gravitational descendants. In some way, and this may be
interesting to investigate further, the infinite brane spectrum we get
here is the boundary analog of the gravitationally dressed chiral ring
$\rnk nk$ of the bulk. This phenomenon may be more general than here in
the context of coset CFT.

\chapter{Final comments}

Having discussed at length the intrinsic algebraic properties of
boundary states of the \nex2 coset models, we may wonder
about the geometrical significance from a space-time, $D$-brane point
of view. There in fact are several ways to associate geometrical data
with the boundary states of the \nex2 coset models, depending on how we
use the BCFT as a building block for constructing a CFT with
geometrical interpretation. For example we can tensor the $A_{k+1}$
minimal model  with a non-compact CFT in order to achieve an integral
central charge $\hat c$, and in this manner one obtains   the
intersection form for the blow-up of $\IC^2/\ZZ_{k+2}$. 
This non-compact geometry, and its
generalization to other \nex2 cosets, is essentially
what we have been discussing so far.

On the other hand, one may build ``Gepner'' tensor products 
of minimal models in order to obtain compact Calabi-Yau manifolds; 
this is the line of research pushed forward
in ref.~\BDLR\ where bundle and sheaf data of these Calabi-Yau's
were determined from the boundary states of the individual component
minimal models. It is an obvious question how this construction
generalizes to tensor products of the more general \nex2 coset models
discussed here; we will present an investigation of this matter
in a future publication. Suffice it to mention here that the
resulting intersection forms and quivers are
closely related to the ones discussed in the present paper,
the main difference being in the multiplicities of the links.

\goodbreak 
\ack
We thank Mike Douglas, Bartomeu Fiol, J\"urg Fr\"ohlich,
Peter Kaste, Andy L\"utken, Peter Mayr, 
Christian R\"omelsberger and Nick Warner 
for valuable discussions, moreover W.L. thanks
the High Energy Group of Rutgers  University for kind hospitality.

\appendix{\appA}{General remarks on the boundary state intersection index}

Assume we start with some rational \nex2\ CFT and a set of
boundary states, with an expansion in terms of Ishibashi
states of the form
$$
|a\rangle = \sum_i \frac{B_{ia}}{\sqrt{S_{i0}}} |i\rangle\rangle \,,
\eqn\bost
$$
where $a$ labels some general boundary condition with some
well-defined automorphism type (A or B-type \OOY) with respect to the \nex2\ 
algebra. Upon inclusion in a string theory, the associated boundary 
states will represent wrapped BPS D-branes.

As in \doubref\DF\HIV, we can define the intersection of two
boundary states, $|a\rangle$ and $|b\rangle$, as an overlap amplitude in the
RR sector. By a modular transformation, this is equal to the Witten
index in the open string Hilbert space on the annulus, with boundary
conditions $a$ and $b$ on the two sides of the annulus, respectively:
$$
\I_{ab} = \langle a|b\rangle_{\rm RR} = tr_{{\cal H}_{ab}}(-1)^F\ .
\eqn\clvsop
$$
The goal of this subsection is to derive a more convenient expression
for $\I_{ab}$, in view of the  cumbersome BCFT expansion \bost. We will
keep the discussion as general  as possible, although in most of the
paper we consider only  ``Cardy'' boundary states.

As a convention, we will assume that the list of Ishibashi
states contains all bosonic-primary fields separately. The
Ishibashi states are normalized as in
$$
\langle\langle i|q^{L_0-c/24}|j\rangle\rangle = \delta_{ij} \chi_i(\tau) \,,
$$
where $q=\ee^{2\pi\ii\tau}$, and $\chi$ is the appropriate 
(bosonic) character.
Upon inserting \bost\ in \clvsop, and recalling that the definition
of the overlap amplitude in the RR sector contains a phase
factor $\ee^{-\pi\ii Q_L(i)}$, we arrive at
$$
\I_{ab} = \sum_{i} \frac{B_{ia}^* B_{ib}}{S_{i0}} 
\ee^{-\pi\ii Q_L(i)} \chi_i(\tau)\ ,
\eqn\inch
$$
where $i$ is summed over all Ishibashi states from the RR sector.
$Q_L(i)$ is the left-moving $U(1)$ charge of the state $i$.
Thus, if $Q_L$ has integer eigenvalues, one can write $\ee^{-\pi\ii 
Q_L}=(-1)^{F_L}$.

The expression \inch\ is in fact independent of $\tau$, and we
can compute it in the limit $\tau\to \ii\infty$, where only Ramond
ground states (Rgs) contribute. Thus,
$$
\I_{ab} = \sum_{i \;{\rm Rgs}} \frac{B_{ia}^* B_{ib}}{S_{i0}} 
\ee^{-\pi\ii Q_i} \,.
\eqn\incl
$$
Let us refer to this expression as the intersection number in the
closed string sector. Several properties of $\I$ can be read off
from \incl. For instance, it is obvious that the rank of $\I$
(viewed as a matrix with entries labelled by the boundary states)
cannot exceed the dimension of the chiral ring (the number
of Ramond ground states is equal to the dimension of the chiral ring).
Therefore, the topological charges of the D-branes lie in a lattice
of rank bounded by the dimension of the chiral ring. What is
not immediate from \incl, however, is the fact that this lattice
is integral. Integrality is more apparent in the open string sector,
as we now demonstrate. Making a modular transformation in \inch,
we obtain
$$
\I_{ab} = \sum_{i, m} \frac{B_{ia}^* B_{ib}S_{im}}{S_{i0}} 
\ee^{-\pi\ii Q_L(i)}
\chi_m(-1/\tau)\ ,
\eqn\inin
$$
where $i$ runs over Ramond Ishibashis and $m$ over all fields. 
We can relax the restriction on $i$ by using that
$$
S_{i, vm} =
\cases{- S_{i,m} & $i$ Ramond sector \cr
S_{i,m} & $i$ Neveu-Schwarz sector}\ ,
$$
where $vm$ denotes the world-sheet superpartner of $m$ ($v$ is
the simple current corresponding to the worldsheet supercurrent).
Furthermore, the $U(1)$ charge is given by half the monodromy
charge with respect to the simple current, $s$, implementing spectral
flow by half a unit. Hence, $\ee^{-\pi\ii Q_L(i)}S_{im}= S_{i,s^{-1}m}$, 
and
$$
\I_{ab} = \frac12 \sum_{i, m}
\frac{B_{ia}^* B_{ib}S_{i,s^{-1}m}}{S_{i0}} (\chi_m(-1/\tau) - \chi_{vm}
(-1/\tau)) 
$$
where now $i$ runs over all fields. We can reduce this expression
by using the well known relation between the Cardy coefficients
and the annulus coefficients, $A_{ab}^{s^{-1}m}=\sum_{i}
B_{ia}^* B_{ib} S_{i,s^{-1}m}/S_{i0}$. The annulus
coefficients are non-negative integers by the Cardy condition.
To obtain a manifestly integral expression, it seems that we should use a 
slightly different normalization for the construction of the true 
supersymmetric boundary states \doubref\BDLR\HIV. Alternatively, the 
factor $1/2$ will be removed in the last steps (GSO projection) of the 
construction of the BPS state. We then write
$$
\I_{ab}
= \sum_m A_{ab}^{s^{-1}m}  (\chi_m(-1/\tau) - \chi_{vm}
(-1/\tau)) \,.
$$
We now recognize $\chi_m-\chi_{vm}$ as a supersymmetric
character. It is equal to one (or $-1$) if $m$ (or $vm$) corresponds
to a Ramond ground states primary and zero otherwise. We then obtain
the intersection number, written in the open string sector with the
help of the annulus coefficients,
$$
\I_{ab} = \sum_{m\; {\rm Rgs}} A_{ab}^{s^{-1}m} - A_{ab}^{vs^{-1}m} \,.
\eqn\inop
$$
The intersection index is now written in a manifestly integer form.  It
follows that the lattice spanned by the boundary states with metric
given by $\I$ is an integral lattice, of rank bounded by the dimension
of the chiral ring.

Various other interesting properties of the intersection
matrix can be derived from \inop\ in a completely model independent
way. For instance, if the \nex2\ theory constitutes the 
internal sector of a string compactification, we have the relation
$sv=c$, where $c$ is the conjugate of the spectral flow.
Using conjugation properties of the annulus coefficients 
and of the chiral ring, one can then show that the intersection 
index is (anti)$^n$-symmetric, where $n$ is  
the number of compact complex dimensions.

\appendix{\appB}{Charge spectrum for models $SU(3)_k/U(2)$}

We here continue the proof that the $\Lambda=0$ states in the standard
range \standardR\ provide an integral basis of the charge lattice, for
the models $SU(3)_k/U(2)$. To this end, we have to express the RR
charges of the boundary states with $\Lambda> 0$ in terms of the basic
($\Lambda=0$) ones. We thus have to find charge vectors
$Q_{(l_1,l_2),\lambda,m}$ with  $(l_1,l_2)$
fixed, that satisfy
$$
\Ih_{(l_1',l_2')(l_1'',l_2'')}
= Q_{(l_1',l_2')}^T \;\; \Ih_{(0,0)(0,0)}
\;\; Q_{(l_1'',l_2'')}\ .
\eqn\Qcond
$$
We claim that the following charge vectors satisfy this
condition. First define
$$
\eqalign{
\tilde Q_{l_1,l_2}
&= N_{l_1} g^{-l_1-2l_2}
+ N_{l_1+1} g^{-l_1-2l_2+3} +
\dots \cr
&\qquad \dots + N_{l_1+l_2} g^{-l_1+l_2}
+ N_{l_1+l_2-1} g^{-l_1+l_2+3} 
+ \dots \cr
 &\qquad\qquad \dots + N_{l_2} g^{l_2+2l_1} \ .
}
$$
Then, if $l_1\ge l_2$
$$
Q_{l_1,l_2} =
\tilde Q_{l_1,l_2}
+ \tilde Q_{l_1-1,l_2-1} 
+\dots 
+ \tilde Q_{l_1-l_2,0}\ ,
\eqn\QttoQ
$$
and the analogous expression if $l_2\ge l_1$.
Indeed, a simple computation shows
$$
\eqalign{
\Ih_{(0,0)(0,0)} \tilde Q_{l_1,l_2}
&= \left(1-N_{k+1}g^{3h}\right)\Bigl[ N_{l_1} g^{-l_1-2l_2}
- N_{l_1+l_2+1}g^{-l_1+l_2+3}  \cr
&\qquad\qquad +N_{l_2}g^{l_2+2l_1+6} 
- N_{l_1-1}g^{-l_1-2l_2+3} \cr
&\qquad\qquad + N_{l_1+l_2-1}g^{-l_1+l_2+3}
-N_{l_2-1}g^{l_2+2l_1+3} \Bigr] \cr
&= \Ih_{(0,0)(l_1,l_2)} - \Ih_{(0,0)(l_1-1,l_2-1)}\ ,
}
$$
where the second term is absent if $l_1=0$ or $l_2=0$.
Thus summing up $\tilde Q$ as in \QttoQ, we obtain,
$$
\Ih_{(0,0)(0,0)} Q_{l_1,l_2}
= \Ih_{(0,0)(l_1,l_2)}\ .
$$
With some more effort, one can check that indeed the 
$Q$'s satisfy \Qcond.

It is quite instructive to draw these charge vectors onto the 
$\Lambda=0$ graph like in \lfig\boundst;
the generalization to more general models becomes then obvious.

\appendix{\appC}{Grassmannian cohomology and principal embeddings of SU(2)}

We review here a concise construction
of the structure constants $\Fp\grl n{k}$ of the
classical cohomology ring $\rnk nk$ \ringiso.
It was used in the physics literature in ref.~\WLwgrav, whose
exposition we follow and where further
details are explained.

It is based on the fact that the ring $\rnk n{k}$ is encoded in the
properties of a particular fundamental representation, $\Xi$, of
$G\equiv SU(\npk)$ (the construction works also for other 
hermitian symmetric spaces $G/H\times U(1)$).  Namely the cohomology
elements are one-to-one to the weights of $\Xi$ and their grade 
($\sim U(1)$ charge) is given, up to a universal shift, by the inner product
of the corresponding weight with the Weyl vector
$\rho\equiv{1\over2}\sum\a_+$.  The highest weight of $\Xi$  (denoted
by $\lambda_\Xi$) is defined by the fundamental weight corresponding to
the node of the $G$-Dynkin diagram that defines the embedding of the
$U(1)$ factor. That is, for the Grassmannians $\gkn n\npk$, $\Xi$ is
given by the $n$-th fundamental representation of $SU(\npk)$ with
highest weight $\lambda_\Xi=(0,\dots,0,1,0,\dots,0)$\ (where $"1"$
appears at the $n$-th entry).

The point is that the matrices $\Fp\grl n{k}$ can be very simply
computed from the weights of the representation $\Xi$ as follows.
Consider the principal $SU(2)$ subgroup of $SU(\npk)$ generated by
$$
\eqalign{
I_+\ &=\
\sum_{{\rm{simple\atop roots\ \alpha}} }a_\alpha^{(1)}\, E_\alpha\cr
I_-\ &=\
\sum_{{\rm{simple\atop roots\ \a}}}a_\alpha^{(-1)} \,E_{-\alpha}\cr
I_0\ &=\ \rho_G\cdot H\ ,}
\eqn\Sdef
$$
where $E_{\pm\a}$ are the generators of $SU(\npk)$ in the
Cartan-Weyl basis and $a_\alpha^{(\pm1)}$ are coefficients such that
$[I_+,I_-]= I_0$. One can always take $a_\alpha^{(1)}\equiv 1$, and
this is what we will assume henceforth. The particular choice of
$I_0$ induces the principal gradation of the generators:
the $I_0$ charge of a generator $E_\a$ is given by $p=\rho\cdot\a$.
The possible values of $|p|$ are just given by 
the exponents $m_i, i=1,\dots,\ell$, of
$G$. One can accordingly group the generators into sets of equal $I_0$ grade, 
and build the following linear combinations:
$$
\eqalign{
\Fp_p\ =\ &\sum_{\{\a:\rho_G\cdot\a=p\}}a_\a^{(p)} E_\a\ ,\ \ \
{\rm\  for\ each\ } p\in\{\pm m_1,\pm m_2,\dots,\pm m_\ell\}\ ,\cr
{\rm with}\ \ &\big[\Fp_p,I_0\,\big]\ =\ p \,\Fp_p\cr}
\eqn\Fpdef
$$
(where $\Fp_1\equiv I_+$, $\Fp_{-1}\equiv I_-$). 
The coefficients $a_\a^{(p)}$ (for
$|p|>1$) are determined from \Sdef\ by requiring:
$$
\big[\Fp_p,\Fp_q\,\big]\ =\ 0 {\rm\ \ \ for\ \ \ }
\cases{ p>0 , q>0\cr p<0 , q<0 \cr}
\eqn\Fpcomm
$$
For $G=SU(\npk)$, one can take $\Fp_p=\sum_{i=1}^{n+k-p}E_{e_i-e_{i+p}}\
(p=1,2,\dots,n+k-1)$.

It is then a crucial fact \KosA\ that, 
when taken in the representation 
$\Xi$ of $G$, the matrices $\Fp_p$ (with $p>0$) generate the 
cohomology ring $\Hst(G/H\times U(1),\IR)$.

In other words, the matrices $\Fp_p, p>0$ represent the LG fields
of the \nex2\ cosets based on $G/H\times U(1)$, the OPE being
represented by simple matrix multiplication. Generic ring elements are
given by polynomials in the $\Fp_p$, the $U(1)$ charge of a ring
element being equal to its $I_0$ grade in units of $1/(g+1)$. In
general, various $\Fp_p$ can be expressed in terms of powers of
lower-degree ${\Fp_q}$ so that they are not independent; which $\Fp_p$
the independent generators are for a given group $G$,  depends on the
representation $\Xi$, \ie, on the choice of $H$. The independent
generators  correspond to a minimal choice of the \LG fields $x_i$.
Obviously, $\Fp_1\equiv I_+$ is always a generator of the ring, and
this corresponds to the fact that in each coset model, there is a
unique \LG field of lowest $U(1)$ charge, $q(x_1)=\coeff1{g+1}$. The
matrices $\Fp_p$ satisfy certain polynomial relations, and these
relations can be integrated to the superpotentials $\wnk nk(x_i)$
defined in \genallW\ (up to simple reparametrizations).

Consider as a first example the \nex2 minimal models of type
$A_{k+1}$, which can be associated to cosets $SU(k\!+\!1)_1/U(k)$,
so that $\Xi$ is the defining representation of $SU(k\!+\!1)$. There is
one independent generator of the chiral ring, which can be
represented by the step generator
$$
\Fp_1\grl1k\ \equiv \ I_+\ =\ \pmatrix
{ 0 & 1 & 0 & \dots & 0 \cr
  0 & 0 & 1 & \dots & 0 \cr
  \vdots & \vdots & \vdots & \ddots &  \vdots \cr
0 & 0 & 0 & \dots & 1 \cr
0 & 0 & 0 & \dots & 0 \cr}_{k+1\times k+1}\ \ ,
\eqn\step
$$
in terms of which the other $\Fp_p$ are given by
$\Fp_p = (\Fp_1)^p$, $p=1,\dots,k$.
The vanishing relation is $(\Fp_1)^{k+1} = 0$,
which corresponds to the \LG\ potential
$\wnk1k(x)=x^{k+2}$.

Next consider the KS model based on $SU(3)_2/U(2)$, which is
equivalent to $SU(4)_1/SU(2)\times SU(2)\times U(1)$. Here 
$\Xi$ is the second, six dimensional fundamental representation with 
$\lambda_\Xi=(0,1,0)$. In this representation we find for
the generators 
\def\aa(#1){\!\!{\fiverm#1}\!}
$$
\Fp_1\grl22\ \equiv \ I_+\ =
\pmatrix{ \aa(0) & \aa(1) & \aa(0) & \aa(0) & \aa(0) & \aa(0) \cr
\aa(0) & \aa(0) & \aa(1) & \aa(    1) & \aa(0) & \aa(0) \cr \aa(0) &
\aa(0) & \aa(0) & \aa(0) & \aa(1) & \aa(0) \cr \aa(    0) & \aa(0) &
\aa(0) & \aa(0) & \aa(1) & \aa(0) \cr \aa(0) & \aa(0) & \aa(0) & \aa(0)
& \aa(    0) & \aa(1) \cr \aa(0) & \aa(0) & \aa(0) & \aa(0) & \aa(0) &
\aa(0) \cr  }\ , \qquad \Fp_2\grl22= \pmatrix{ \aa(0) & \aa(0) &
\aa(0) & \aa(1) & \aa(0) & \aa(0) \cr \aa(0) & \aa(0) & \aa(    0) &
\aa(0) & \aa(1) & \aa(0) \cr \aa(0) & \aa(0) & \aa(0) & \aa(0) & \aa(0)
& \aa(    0) \cr \aa(0) & \aa(0) & \aa(0) & \aa(0) & \aa(0) & \aa(1)
\cr \aa(0) & \aa(0) & \aa(    0) & \aa(0) & \aa(0) & \aa(0) \cr \aa(0)
& \aa(0) & \aa(0) & \aa(0) & \aa(0) & \aa(    0) \cr  } \ ,
\eqn\Fsix
$$
where we have made a change of basis, ie., $\Fp_2\to1/2(\Fp_2+(\Fp_1)^2)$.
These matrices represent the LG fields $x_1$, $x_2$, resp.,
of the superpotential 
$W\grl22={1\over5}x_1^5 - {{x_1}^3}\,x_2 + x_1\,{{x_2}^2}$,
which describes a minimal model of type $D_6$.

Finally consider the KS model based on $SU(3)_3/U(1)$, which is
equivalent to $SU(5)_1/SU(2)\times SU(3)\times U(1)$. Here 
$\Xi$ is the second, ten dimensional fundamental representation with 
$\lambda_\Xi=(0,1,0,0)$. In this representation we find for
the generators 
$$
\Fp_1\grl23\ \equiv \ I_+\ =
\pmatrix{ \aa(0) & \aa(1) & \aa(0) & \aa(0) & \aa(0) & \aa(0) & \aa(0)
& \aa(0) & \aa(0) & \aa(    0) \cr \aa(0) & \aa(0) & \aa(1) & \aa(0) &
\aa(1) & \aa(0) & \aa(0) & \aa(0) & \aa(0) & \aa(    0) \cr \aa(0) &
\aa(0) & \aa(0) & \aa(1) & \aa(0) & \aa(1) & \aa(0) & \aa(0) & \aa(0) &
\aa(    0) \cr \aa(0) & \aa(0) & \aa(0) & \aa(0) & \aa(0) & \aa(0) &
\aa(1) & \aa(0) & \aa(0) & \aa(    0) \cr \aa(0) & \aa(0) & \aa(0) &
\aa(0) & \aa(0) & \aa(1) & \aa(0) & \aa(0) & \aa(0) & \aa(    0) \cr
\aa(0) & \aa(0) & \aa(0) & \aa(0) & \aa(0) & \aa(0) & \aa(1) & \aa(1) &
\aa(0) & \aa(    0) \cr \aa(0) & \aa(0) & \aa(0) & \aa(0) & \aa(0) &
\aa(0) & \aa(0) & \aa(0) & \aa(1) & \aa(    0) \cr \aa(0) & \aa(0) &
\aa(0) & \aa(0) & \aa(0) & \aa(0) & \aa(0) & \aa(0) & \aa(1) & \aa(
0) \cr \aa(0) & \aa(0) & \aa(0) & \aa(0) & \aa(0) & \aa(0) & \aa(0) &
\aa(0) & \aa(0) & \aa(    1) \cr \aa(0) & \aa(0) & \aa(0) & \aa(0) &
\aa(0) & \aa(0) & \aa(0) & \aa(0) & \aa(0) & \aa(    0) \cr  }\ ,\qquad
\Fp_2\grl23= \pmatrix{ \aa(0) & \aa(0) & \aa(0) & \aa(0) & \aa(1) &
\aa(0) & \aa(0) & \aa(0) & \aa(0) & \aa(    0) \cr \aa(0) & \aa(0) &
\aa(0) & \aa(0) & \aa(0) & \aa(1) & \aa(0) & \aa(0) & \aa(0) & \aa(
0) \cr \aa(0) & \aa(0) & \aa(0) & \aa(0) & \aa(0) & \aa(0) & \aa(1) &
\aa(0) & \aa(0) & \aa(    0) \cr \aa(0) & \aa(0) & \aa(0) & \aa(0) &
\aa(0) & \aa(0) & \aa(0) & \aa(0) & \aa(0) & \aa(    0) \cr \aa(0) &
\aa(0) & \aa(0) & \aa(0) & \aa(0) & \aa(0) & \aa(0) & \aa(1) & \aa(0) &
\aa(    0) \cr \aa(0) & \aa(0) & \aa(0) & \aa(0) & \aa(0) & \aa(0) &
\aa(0) & \aa(0) & \aa(1) & \aa(    0) \cr \aa(0) & \aa(0) & \aa(0) &
\aa(0) & \aa(0) & \aa(0) & \aa(0) & \aa(0) & \aa(0) & \aa(    0) \cr
\aa(0) & \aa(0) & \aa(0) & \aa(0) & \aa(0) & \aa(0) & \aa(0) & \aa(0) &
\aa(0) & \aa(    1) \cr \aa(0) & \aa(0) & \aa(0) & \aa(0) & \aa(0) &
\aa(0) & \aa(0) & \aa(0) & \aa(0) & \aa(    0) \cr \aa(0) & \aa(0) &
\aa(0) & \aa(0) & \aa(0) & \aa(0) & \aa(0) & \aa(0) & \aa(0) & \aa(
0) \cr  }\ ,
$$
where we have made the same reparametrization as above.
These matrices correspond to the superpotential
$W\grl 23= {1\over6}{x_1}^6 - {{x_1}^4}\,x_2 + {3\over2}
{{x_1}^2}\,{{x_2}^2} - {1\over3}{{x_2}^3}$,
which is of singularity type $J_{10}$ \arn. 
The deformed matrices
$$
\hat\Fp_1\grl22(\mu)=\Fp_1\grl23+\mu \Fp_{-4}\grl23=
\pmatrix{ \aa(0) & \aa(1) & \aa(0) & \aa(0) & \aa(0) & \aa(0) & \aa(0)
& \aa(0) & \aa(0) & \aa(    0) \cr \aa(0) & \aa(0) & \aa(1) & \aa(0) &
\aa(1) & \aa(0) & \aa(0) & \aa(0) & \aa(0) & \aa(    0) \cr \aa(0) &
\aa(0) & \aa(0) & \aa(1) & \aa(0) & \aa(1) & \aa(0) & \aa(0) & \aa(0) &
\aa(    0) \cr \aa(0) & \aa(0) & \aa(0) & \aa(0) & \aa(0) & \aa(0) &
\aa(1) & \aa(0) & \aa(0) & \aa(    0) \cr \aa(0) & \aa(0) & \aa(0) &
\aa(0) & \aa(0) & \aa(1) & \aa(0) & \aa(0) & \aa(0) & \aa(    0) \cr
\aa(0) & \aa(0) & \aa(0) & \aa(0) & \aa(0) & \aa(0) & \aa(1) & \aa(1) &
\aa(0) & \aa(    0) \cr \aa(\mu ) & \aa(0) & \aa(0) & \aa(0) & \aa(0) &
\aa(0) & \aa(0) & \aa(0) & \aa(    1) & \aa(0) \cr \aa(0) & \aa(0) &
\aa(0) & \aa(0) & \aa(0) & \aa(0) & \aa(0) & \aa(0) & \aa(    1) &
\aa(0) \cr \aa(0) & \aa(\mu ) & \aa(0) & \aa(0) & \aa(0) & \aa(0) &
\aa(0) & \aa(    0) & \aa(0) & \aa(1) \cr \aa(0) & \aa(0) & \aa(\mu ) &
\aa(0) & \aa(0) & \aa(0) & \aa(    0) & \aa(0) & \aa(0) & \aa(0) \cr  }
$$
and
$$
\hat\Fp_2\grl22(\mu)=\Fp_2\grl23+\mu \Fp_{-3}\grl23=
\pmatrix{ \aa(0) & \aa(0) & \aa(0) & \aa(0) & \aa(1) & \aa(0) & \aa(0)
& \aa(0) & \aa(0) & \aa(    0) \cr \aa(0) & \aa(0) & \aa(0) & \aa(0) &
\aa(0) & \aa(1) & \aa(0) & \aa(0) & \aa(0) & \aa(    0) \cr \aa(0) &
\aa(0) & \aa(0) & \aa(0) & \aa(0) & \aa(0) & \aa(1) & \aa(0) & \aa(0) &
\aa(    0) \cr \aa(\mu ) & \aa(0) & \aa(0) & \aa(0) & \aa(0) & \aa(0) &
\aa(0) & \aa(0) & \aa(    0) & \aa(0) \cr \aa(0) & \aa(0) & \aa(0) &
\aa(0) & \aa(0) & \aa(0) & \aa(0) & \aa(1) & \aa(    0) & \aa(0) \cr
\aa(0) & \aa(0) & \aa(0) & \aa(0) & \aa(0) & \aa(0) & \aa(0) & \aa(0) &
\aa(    1) & \aa(0) \cr \aa(0) & \aa(\mu ) & \aa(0) & \aa(0) & \aa(0) &
\aa(0) & \aa(0) & \aa(    0) & \aa(0) & \aa(0) \cr \aa(0) & \aa(0) &
\aa(0) & \aa(0) & \aa(0) & \aa(0) & \aa(0) & \aa(    0) & \aa(0) &
\aa(1) \cr \aa(0) & \aa(0) & \aa(\mu ) & \aa(0) & \aa(0) & \aa(0) &
\aa(    0) & \aa(0) & \aa(0) & \aa(0) \cr \aa(0) & \aa(0) & \aa(0) &
\aa(\mu ) & \aa(0) & \aa(    0) & \aa(0) & \aa(0) & \aa(0) & \aa(0) \cr
 }\ ,
$$
then represent the LG fields of the perturbed superpotential
$\hat W\grl 22(x_i,\mu)\equiv W\grl 23(x_i)+\mu x_1$.
They play a r\^ole for the extended
intersection form $\Ih\grl 22$ in \Itwotwo, when
specialized to $\mu=-1$. When specialized to $\mu=+1$, they
are the fusion coefficients of $U(2)_{k+1=2,h=5}$.

\goodbreak
\refout
\end